\documentclass{iopart}

\expandafter\let\csname equation*\endcsname\relax
\expandafter\let\csname endequation*\endcsname\relax

\usepackage{lipsum} 
\usepackage{color}
\usepackage{graphicx}
\usepackage{float}
\usepackage{iopams}
\usepackage[margin=10pt,font=small,format=hang]{caption}
\usepackage{amsmath,amssymb}
\usepackage{acronym}
\usepackage{multirow}
\usepackage{tabu}
\usepackage{tikz}
\usetikzlibrary{arrows,shapes,trees,decorations.pathreplacing}
\usepackage{import}
\usepackage{svn-multi}
\usepackage{lineno}
\usepackage{hyperref}
\usepackage{mathtools}
\usepackage{xspace}
\usepackage{tabularx}
\usepackage{url}
\usepackage{cite}
\usepackage{lineno}
\usepackage[normalem]{ulem}
\hypersetup{
    colorlinks=true,
    linkcolor=blue,
    filecolor=magenta,      
    urlcolor=cyan,
}

\DeclareGraphicsExtensions{%
    .pdf,.PDF,%
    .png,.PNG,%
    .jpg,.mps,.jpeg,.jbig2,.jb2,.JPG,.JPEG,.JBIG2,.JB2}

\newcommand{\FP}{Fabry-P\'{e}rot\xspace}
\newcommand{\CE}{Cosmic Explorer\xspace}

\usepackage{color}

\makeatletter
\newcommand\footnoteref[1]{\protected@xdef\@thefnmark{\ref{#1}}\@footnotemark}
\makeatother

\begin{document}

\title[Technical Noise, Data Quality, and Calibration Requirements for XG]{Technical Noise, Data Quality, and Calibration Requirements for Next-Generation Gravitational-Wave Science}
\author{E Capote$^1$,
        L Dartez$^2$,
        D Davis$^3$
        }

\address{$^1$Syracuse University, Syracuse, NY 13244, USA}
\address{$^2$LIGO Hanford Observatory, Richland, WA 99352, USA}
\address{$^3$LIGO, California Institute of Technology, Pasadena, CA 91125, USA}
\date{\today}

\begin{abstract}

The next generation of ground-based gravitational-wave interferometers is expected to generate a bounty of new astrophysical discoveries, with sensitivities and bandwidths greatly improved compared to current-generation detectors. 
These detectors will allow us to make exceptional advancements in our understanding of fundamental physics, the dynamics of dense matter, and the cosmic history of compact objects. 
The fundamental design aspects of these planned interferometers will enable these new discoveries; 
however, challenges in technical noise, data quality, and calibration have the potential to limit the scientific reach of these instruments.
In this work, we evaluate the requirements of these elements for next-generation gravitational-wave science, focusing on how these areas may impact the proposed Cosmic Explorer observatory.
We highlight multiple aspects of these fields where additional research and development is required to ensure Cosmic Explorer reaches its full potential.
\end{abstract}

\maketitle

\section{Introduction}\label{sec:intro}

Advances in gravitational-wave detector technology have pushed the limits of possibility for the detection of new and exciting gravitational wave events~\cite{GW150914_paper,GW170817,GWTC-1,GWTC-2,GWTC-2.1,GWTC-3}. Despite many advances, ground-based gravitational-wave detectors such as Advanced LIGO~\cite{aligo}, Advanced Virgo~\cite{avirgo}, and KAGRA~\cite{kagra} continue to be limited in sensitivity and operation due to both fundamental limits of gravitational-wave detectors and technical challenges in their operation~\cite{o3LIGOinstrumentation,Virgo_O3_detchar,KAGRA_O3GK_performance,KAGRA_detchar}.

Several next-generation (XG) gravitational-wave detectors have been proposed, such as Cosmic Explorer~\cite{CE_horizon}, planned to be located in the United States, and the Einstein Telescope~\cite{ET_design}, planned to be located in Europe. Designs for these new detectors are based on current understanding of the best technologies for gravitational-wave detector performance. To enhance these new detectors, several key upgrades are proposed that push the limits of current technology to further improve detector sensitivity.

Noise sources that limit these types of gravitational-wave detectors are usually divided into a few main categories. Fundamental noise sources are derived from first principles and set an upper limit to any detector sensitivity.
Baseline designs for XG detectors tackle many of these fundamental sources first by lengthening the \FP{} cavity arms and increasing the mass and size of the test masses. Additionally, these designs include enhanced seismic isolation requirements through longer and larger suspensions, improved thermal noise via lower loss optical coatings, and reduction in quantum noise through high optical power and  high levels of quantum squeezing~\cite{CE_horizon,ET_design,NEMO}.
Meanwhile, technical noises result from the operation of a detector and can usually be reduced if studied and understood. While baseline design choices focus on fundamental noise reduction, some design factors also tackle technical noise challenges. Some work has already been done to consider the design of the \CE detector with knowledge of fundamental, technical, and environmental limiting noises at low frequency, as in Ref.~\cite{CE_LF}.

Technical noises are typically considered to be stationary but can also be non-stationary over long time periods or even create short bursts of transient noise called ``glitches''~\cite{GW150914_detchar,Nuttall:2018xhi,LIGO_O3_detchar,Virgo_O3_detchar,KAGRA_detchar}. 
Technical noise effects that are non-stationary or transient are often considered separately and are referred to as ``data quality issues.'' 
While the origin of these time-varying technical noises is similar to other sources of technical noise, they can create additional challenges for data analysis pipelines compared to stationary noise~\cite{LIGOScientific:2019hgc,Davis:2022dnd,TheLIGOScientific:2017lwt,LIGO_O3_detchar,Powell:2018csz,Kwok:2021zny,Mozzon:2021wam,Macas:2022afm,Hourihane:2022doe,Ghonge:2023ksb,Davis:2022ird}.
For this reason, understanding and mitigating data quality issues is an active area of research and vitally important to improving the robustness of astrophysical analyses with gravitational waves~\cite{Nuttall:2018xhi,2023ApPhL.122r4101B,Davis:2022dnd}.

Alongside technical noise and data quality, an additional important
consideration is the calibration of the detector data. Calibration is the
process of converting raw detector signals into measures of gravitational-wave
strain experienced by the detector~\cite{Sun_2020,Abbott_2017}. The uncertainty in the detector's
calibration has the potential to directly impact the measurability of
astrophysical parameters~\cite{Essick_2022,Sun_2020,Cahillane:2017vkb,Virgo:Cal_O2_Acernese_2018,Karki_2016,Bhattacharjee_2020,galaxies10020042}. Given the anticipated sensitivity of these
detectors, calibration uncertainty has the potential to be a limiting factor in
astrophysical analyses that use XG data. As such, it is imperative for the
operation of future gravitational-wave detectors that calibration uncertainty
and systematic error therein are mitigated, understood, and quantified with
care.

Addressing these challenges will require active research that builds upon current work being done to design XG detectors.  
In this work, we focus on \CE as an example of an XG gravitational-wave detector to explore the impacts of technical noise, data quality, and calibration on XG gravitational-wave science.
We also compare and contrast how these issues may affect the Einstein Telescope when appropriate. \CE is planned to be a dual-recycled Michelson interferometer with 40\,km or 20\,km Fabry-P\'{e}rot arms~\cite{CE_horizon}. The test masses serving as the arm cavity mirrors are planned to be 320\,kg~\cite{CE_horizon}, eight times the mass of the current Advanced LIGO test masses~\cite{aligo}. These and other design choices, such as improved seismic isolation, will allow \CE to achieve unprecedented sensitivity to gravitational waves.
Due to the focus on \CE, the projections included in this work use previous results from the Advanced LIGO (aLIGO) detectors, as these are the current-generation detectors most similar to the proposed \CE design~\cite{CE_horizon}.

First, we consider technical noises that create broadband, stationary noise in the detector, and possible mitigation strategies (Section~\ref{sec:tech_noise}). 
Next, we consider the effect of transient and non-stationary noise, especially with regard to data quality (Section~\ref{sec:trans_noise}). 
We then discuss the challenges of narrow-band technical noises (Section~\ref{sec:lines}) 
and technical noises correlated across detector networks (Section~\ref{sec:corr_noise}).
We also consider the calibration efficacy in XG detectors (Section~\ref{sec:cal}).
Finally, we comment on the outlook of addressing these challenges to best realize XG science goals (Section~\ref{sec:discuss}).

\section{Broadband Technical Noises}\label{sec:tech_noise}

To begin the discussion of technical noises and their impacts, we focus on technical noises that impact the detector over a broad range of frequencies.
In current generation detectors, one of the most significant sources of technical noise, especially at low frequency, comes from auxiliary controls such as length and alignment controls~\cite{o1detectorpaper, o3LIGOinstrumentation,Virgo_O3_detchar,KAGRA:2022fgc}. Control servos are a necessary part of sensitive interferometer operation to achieve stable resonance conditions in optical cavities, ensure good optical gain, and maintain operation in the linear control regime. However, controls can also inject excess noise into the gravitational wave band through various means. Part of the operation of a gravitational-wave interferometer involves finding a delicate balance between an optimal control point and minimal noise as these are two situations that can be at odds~\cite{o3LIGOinstrumentation}. In this discussion we will enumerate controls requirements as they are in current generation detectors, and what considerations are required for next generation science goals.

\subsection{Auxiliary Length Control Noise}\label{sec:lsc}

Auxiliary cavity length control is an essential part of stable interferometer operation. Assuming \CE continues to follow a similar topology as current advanced gravitational-wave detectors, these auxiliary lengths are the power recycling cavity (PRC) length, the Michelson cavity (MICH) length, and signal extraction cavity (SEC) length~\cite{CE_LF}. Auxiliary lengths couple marginally to the differential arm length (DARM) and, therefore, noise related to controlling these cavities couples into the DARM spectrum from which the gravitational-wave readout is derived~\cite{o3LIGOinstrumentation}. The significant noise sources in aLIGO are from the MICH length and the SEC\footnote{In many aLIGO papers this is referred to as the ``signal recycling cavity'', or ``SRC''.} length length~\cite{o3LIGOinstrumentation}. Advanced LIGO does not experience any significant noise contribution from the PRC length, and this is so far expected to be the same for \CE. In principle, no noise will couple from the SEC in \CE due to the balanced homodyne readout~\cite{CE_LF}, however, in practice effects like contrast between each arm cavity mode could result in SEC noise coupling. For the purpose of this paper, we assume that the main auxiliary noise coupling results from MICH.

Michelson cavity motion creates differential phase sidebands that become amplitude sidebands at the output of the interferometer.
However, this noise is suppressed in the interferometer via the arm cavity finesse, as the DARM signal is amplified by the \FP arm gain relative to the MICH length signal.
Current design considerations of \CE assume the same finesse as aLIGO~\cite{CE_horizon}.
Therefore, the Michelson coupling will have approximately the same factor, $\pi/2\mathcal{F} = 1/280$.
 However, studies are underway to consider the benefits of changing the \CE finesse; see Ref.~\cite {Srivastava_2022} for a more in-depth discussion.

To understand the effects of noise coupled from Michelson length control, it is
beneficial to understand what noises limit the Michelson loop itself. In Advanced LIGO, the Michelson loop was shown to be
limited by sensing noise roughly above 10\,Hz, at about $10^{-16}$
m$/\sqrt{\mathrm{Hz}}$. This sensing noise is shown in Fig.~9a of Ref.~\cite{o1detectorpaper} to result from the quantum (shot) noise on the photodetector used to detect the MICH length signal. Below 10\,Hz, MICH is limited by suspension damping noise from the local control of the
beamsplitter, the mirror used to control Michelson
length~\cite{o1detectorpaper}. Besides suspension damping noise, there is also seismic noise, roughly following a $1/f^6$ slope and reaches a level of 1$\times10^{-17}$ m/$\sqrt{\mathrm{Hz}}$ at 10\,Hz. The slope of this noise results from the triple suspension used to suspend the beamsplitter.
All of these noises couple to DARM and limit the gravitational-wave channel, though for the aLIGO band which extends only to 10\,Hz, the most significant noise is MICH sensing noise.

To estimate of the level of MICH noise in a gravitational-wave detector, we can take the current level of MICH sensing noise above 10\,Hz, $10^{-16}$ m$/\sqrt{\mathrm{Hz}}$, and multiply it with the assumed coupling factor, $1/280$. The resulting displacement noise is approximately 3$\times 10^{-19}$ m$/\sqrt{\mathrm{Hz}}$. This noise level is significant, and completely limits Advanced LIGO if left unmitigated. However, a feedforward control is applied that suppresses this noise out of band. Feedforward is a form of real-time noise cancellation based on a known transfer function of MICH to DARM coupling~\cite{aligo,Meadors:2013lja}. Currently, it is possible to achieve at least a factor of one hundred suppression of this sensing noise in the gravitational-wave band. Such a suppression factor is dependent on the measurement---a measurement precision of 1\% is required.

One drawback of feedforward is that it can be tuned to suppress sensing noise or displacement noise, but not both.
Given the high level of MICH sensing noise that can couple to the gravitational-wave band, feedforward subtraction must be tuned to reduce sensing noise, but will also, by the nature of feedforward, amplify any displacement noise present.
In Advanced LIGO, the bandwidth of the MICH loop is set to around 10\,Hz. There are several considerations for the setpoint of the MICH bandwidth, but the one most important for this discussion is the suppression of displacement noise. In order to maintain the linear control regime required for detector operation, the residual displacement of MICH length must be around a few picometers~\cite{o1detectorpaper, o3LIGOinstrumentation}. This bandwidth is high enough to achieve the required displacement noise residual for aLIGO and any amplified displacement noise that occurs is below the observation band for gravitational waves.

However, the \CE band will extend down to 5\,Hz, so the effect of MICH displacement noise must also be considered.
Part of the design for \CE involves better local control for suspensions, namely, making use of local sensors with much lower noise~\cite{CE_LF,CE_horizon}. The proposed Homodyne Quadrature Interferometer (HoQI), is one such sensor that has the required performance for a detector like \CE~\cite{CooperHoQI}. The benefit of using the HoQI would be much less suspension damping noise, such that the low frequency displacement noise that limits MICH would likely be limited by seismic noise. To further improve upon displacement noise, \CE design also considers improved seismic isolation via improved suspension design, such that \CE could achieve a factor of ten reduction in seismic noise at 1\,Hz~\cite{CE_LF}. These improvements allow for the ability to relax the controls design for MICH. Feedforward will still be required for \CE: in a 40\,km detector, a MICH loop with sensing noise of $10^{-16}$ m$/\sqrt{\mathrm{Hz}}$ and an arm finesse of 450 will result in a strain noise of about 6$\times 10^{-24}$ 1$/\sqrt{\mathrm{Hz}}$~\cite{CE_LF}. If a one hundred times suppression of the MICH sensing noise is achieved, this noise would be at least five times below the level of the \CE fundamental noise curve, which reaches 3$\times 10^{-25}$ 1$/\sqrt{\mathrm{Hz}}$ at 20\,Hz~\cite{CE_LF,o3LIGOinstrumentation}. However, any residual displacement noise below 10\,Hz will be amplified by this feedforward, which could completely contaminate the detection band from 5-10\,Hz.

By suppressing MICH displacement noise with better local controls and seismic isolation, the \CE MICH bandwidth could be set below 5\,Hz. With less MICH loop gain above 5\,Hz, less sensing noise will be coupled in the band, and less displacement noise will be amplified by the feedforward. This has two beneficial impacts: more sensitivity in the region from 5-10\,Hz, and less stringent requirements on the feedforward suppression. To completely avoid any risk of displacement noise contaminating the gravitational-wave strain at 5\,Hz, a MICH bandwidth that is closer to 3\,Hz would be the most beneficial. Loop bandwidth cannot be lowered indefinitely, both for displacement noise considerations, and for control of any suspension resonances. It will be key to keep the beamsplitter suspension length resonances low enough that such a loop design is possible for \CE.

Assuming no change in the sensing noise level, that means the seismic noise needs to be less than $10^{-16}$\,m$/\sqrt{\mathrm{Hz}}$ below 5\,Hz. Seismic noise reaches approximately $5\times10^{-15}$\,m$/\sqrt{\mathrm{Hz}}$ at 5\,Hz~\cite{o1detectorpaper} in aLIGO. 
Switching to an improved suspension design (see ~\cite{CE_LF}), and changing the beamsplitter suspension from triple in aLIGO to a quadruple suspension to gain an additional additional $1/f^{2}$ suppression, can reduce the seismic noise at 5\,Hz below the sensing noise as required. This is one path forward to keep seismic noise low enough that any amplification of displacement noise from the feedforward would be kept out of the \CE band. Overall, even with little change in the MICH sensing noise, design strategies related to reducing displacement noise via improved local damping and improved seismic isolation will have a significant effect on suppressing any length controls noise that can contaminate above 5\,Hz.

These projections are designed to be only a first look at potential auxiliary length coupling in XG detectors. As the designs for \CE and other XG detectors develop, simulation can further investigate how this noise may appear in an XG detector. Elements such as seismic isolation, suspension design, and optical layout could affect the appearance of Michelson noise or even other auxiliary lengths. Notably missing here is a full corner optical layout for the \CE interferometer, showing the planned lengths of the auxiliary cavities, and the mirrors and telescopes required. Development of such a layout is still underway as a part of the \CE preliminary design work, and better modeling of auxiliary length noise can be undertaken once this layout is proposed.
Furthermore, any changes to the \FP cavity finesse will adjust the MICH coupling; increasing the finesse will be beneficial in reducing MICH noise while lowering the finesse benefits SEC losses that limit high-frequency sensitivity~\cite{Srivastava_2022}.
Even in current detectors, there is still much to be studied about auxiliary length coupling. For example, the projections here only consider a linear coupling of MICH to DARM, but some evidence from O3 indicates that auxiliary length coupling could have a nonlinear component~\cite{o3LIGOinstrumentation}. Nonetheless, this initial work highlights the importance of studying this type of technical noise as a part of maximizing XG detector sensitivity.

\subsection{Alignment Sensing and Control Noise}\label{sec:asc}

Alignment sensing and control, or ASC, is arguably one of the most topologically complex controls system in ground-based interferometers and is often a limiting noise at low frequency~\cite{o3LIGOinstrumentation}. The alignment of the \FP arm cavities and the auxiliary cavities must be controlled together: one nanoradian residual angular motion is required for the arms, while tens of nanoradians of residual motion is acceptable for auxiliary cavities~\cite{BarsottiASC}. These requirements are set to stabilize resonant conditions, suppress higher order spatial modes that can cause optical loss, and reduce cross coupling between degrees of freedom.

In aLIGO, alignment degrees of freedom are sensed via quadrant photodiodes (QPDs) and wavefront sensors (WFS)~\cite{BarsottiASC}. QPDs provide pointing control for while WFS are interferometric sensors using radio frequency quadrant photodetectors to sense the relative position of the carrier and sidebands within the cavities~\cite{o3LIGOinstrumentation}. A third type of alignment control, called dither control, will be briefly discussed in Section~\ref{sec:lines}. Beams are picked off at different ports of the interferometer to provide signals for these sensors.
It is likely similar alignment sensing schemes and sensor layouts will be applied in \CE. Given the residual angular motion requirements, the alignment controls that provide the most significant noise impacts are the arm controls--- achieving one nanoradian RMS requires loop bandwidths around a few Hz. Control of auxiliary cavity alignments, such as the PRC, MICH, and SEC alignments, are important, but the bandwidth required for these loops in Advanced LIGO is less than 1\,Hz to achieve the RMS requirement. Therefore, this paper will focus mainly on the control of the arms when considering the technical noise.

The \FP arm control is performed in the cavity basis, known as ``hard" and ``soft" modes (the reason for these names will be explained shortly). A hard mode results from a shift in the cavity axis alignment, while the soft mode results in a translation of the beam position within the cavity. The hard mode is typically detected on a wavefront sensor (WFS), which detects the first-order misalignment mode from a cavity axis shift,
while the soft mode is sensed at a quadrant photodiode (QPD) after the beam reduction telescope. In Advanced LIGO, the highest bandwidth control in alignment sensing is from the hard loops.
Above 10\,Hz, the limiting noise source for both WFS and QPDs is the sensing noise of the detector, about $10^{-14}$ rad$/\sqrt{\mathrm{Hz}}$~\cite{BarsottiASC}. However, in aLIGO, some of these sensors are contaminated with excess noise from 10-30\,Hz, due to spurious vertical and rotational coupling of the motion of the in-vacuum table where the sensors are mounted~\cite{o3LIGOinstrumentation}. This particular excess noise contamination is the result of poor seismic isolation of the table itself and can elevate the sensor noise level in the WFS, magnifying the injected noise from ASC into the DARM spectrum. While there have been some successful mitigation strategies using feedforward subtraction or signal blending in aLIGO~\cite{o3LIGOinstrumentation}, next-generation detector design should consider table isolation of similar importance to mirror isolation when it comes to reducing technical noise from alignment controls.

The coupling of angular mirror motion to cavity length is best understood via a geometric argument as
\begin{equation}
    \Delta L = d_{spot} \times \theta_{mirror},
    \label{eq:asc_coupling}
\end{equation}
where $d_{spot}$ represents the beam spot motion on the mirror, and $\theta_{mirror}$ represents the mirror's residual angular motion~\cite{BarsottiASC}. This is a bilinear process, as both beam spot motion and mirror angular motion can have a separate frequency dependence. By achieving the nanoradian RMS requirement in the arm cavities, for example, the effect of the beam spot motion that is convolved with the residual angular motion is suppressed and has minimal noise coupling. In this case, the remaining noise coupling is linear, and results from residual angular motion coupling to a static beam offset~\cite{BarsottiASC}.
The coupling of the arm cavity mirrors to the differential arm length, DARM, is the strongest, as any change in the angle of an arm cavity mirror is a direct change in DARM. The auxiliary cavity mirrors also experience a coupling to DARM, usually with a coupling strength of two or more orders of magnitude below the coupling strength of the arm cavity mirrors~\cite{BarsottiASC}. This is further motivation to consider only the noise coupling of the arm degrees of freedom as the dominant noise source from ASC.

Assuming a similar ASC sensing noise level as aLIGO, 10$^{-14}$ rad$/\sqrt{\mathrm{Hz}}$, and a similar residual beam spot motion of 0.1\,mm~\cite{BarsottiASC}, the estimated strain noise from ASC at 20\,Hz in \CE would be about 2.5$\times10^{-23}$ 1$/\sqrt{\mathrm{Hz}}$. In order to suppress this noise in the gravitational-wave band, low pass filtering can be applied to the alignment control loop. Considering the arm alignment loop bandwidth in aLIGO~\cite{o3LIGOinstrumentation}, two orders of magnitude loop suppression is achievable by 20\,Hz, which would place the resulting strain noise from ASC just below the \CE design sensitivity at 20\,Hz. However, this is not sufficient to achieve the desired \CE sensitivity at 5\,Hz. If the alignment control loop bandwidth can be reduced, the sensing noise can be suppressed more in the detection band. This requires a reduction of the angular displacement noise, such that less control is required to meet the same residual motion requirements. Loop bandwidth reduction will also depend on the suspension design, as the suspension dynamics play a significant part of the design of the arm alignment control. In particular, the optical torque from the high power in the arms will affect the angular suspension dynamics.

The high circulating power in the arm cavities creates a radiation pressure which exerts a torque to the suspended mirrors. This torque provides an additional mechanical term to the suspension system, which ``hardens'' (mirrors rotate with the same sign) or ``softens'' (mirrors rotate with opposite sign) the suspension modes~\cite{SidlesSigg}. Given the high expected operating power of \CE (1.5\,MW)~\cite{CE_horizon}, this radiation pressure effect will be significant for determining the success of the interferometer alignment stability and resulting noise couplings.

The hard and soft suspension modes will be shifted in frequency from the free suspension resonance, $f_0$, via the factor~\cite{CE_LF,BarsottiASC,SidlesSigg}
\begin{equation}
    \Delta f_{h,s}^2 = \frac{\gamma_{h,s} P_{cav}L_{arm}}{cI_{m}},
    \label{eq:asc_freq}
\end{equation}
where $\gamma_{h,s}$ are the geometric factors of the hard/soft mode, $P_{cav}$ and $L_{arm}$ represent the intra-cavity power and arm length, respectively, and $I_{m}$ and $c$ represent the mirror moment of inertia and speed of light. The factor $\gamma_{h,s}$ depends only on the cavity geometry (length of the arm and radii of curvature of the test masses);
$\gamma_h > 0$, as the hard mode increases in frequency (stiffens) with increasing cavity power, and $\gamma_s < 0$, the soft mode reduces in frequency (softens) with increasing cavity power. If the soft mode frequency shift, $\Delta f_{s}^2$, exceeds the fundamental frequency of the suspension, $f_0$, the soft mode becomes completely unstable. Although the hard mode frequency shift is always positive and always stable, the suspension mode must be damped~\cite{CE_LF}. As such, the hard mode frequency shift, $\Delta f_{h}^2$, sets a lower limit for the bandwidth of any alignment control loop applied to the arm alignment control.

With these dynamics in mind, much of the consideration for the technical noise impact of alignment controls noise comes directly from the chosen design of the suspension. Current \CE includes a quadruple pendulum system similar to aLIGO's, scaled up to support the 320\,kg test masses~\cite{CE_horizon}. A suspension design with low-frequency mechanical resonances will be required to keep the hard mode frequency low enough to design a low bandwidth control loop. Conversely, it would be beneficial to maintain a high enough suspension frequency that the soft mode does not become unstable. With the intra-cavity power and arm length for \CE set to be 1.5\,MW and 40\,km respectively, adjusting the cavity geometry also allows some control over the impact of the radiation pressure effect. It is beneficial for \CE design to limit diffraction over 40\,km, so the design keeps the beam size as small as possible. This design component has an added benefit for alignment control considerations, as smaller beams have smaller geometric factors, shifting the suspension resonances less~\cite{SidlesSigg}.
Furthermore, it is beneficial to increase the mirror moment of inertia, $I_{m}$, as this would also reduce the shift in frequency from high power. If the mass of the mirror is fixed, this involves adjusting the aspect ratio of the mirror to maximize $I_{m}$.

For the current parameters in the \CE design, the resulting frequency shift of the hard mode is likely to be around $+$(1.1\,Hz)$^2$ and for the soft mode $-$(0.6\,Hz)$^2$, as found in Ref.~\cite{CE_LF}.
Both the hard and soft modes should be considered in suspension design; minimizing the instability of the soft mode is beneficial, as a control bandwidth for an unstable soft mode will need to be around $3f_s$~\cite{CE_LF}. The bandwidth of control for the hard mode will also determine how much the injected angular noise can be suppressed in the gravitational-wave band---the lower the bandwidth, the more suppression at 5\,Hz is possible. The quadruple suspension design will involve four main eigenmodes of angular motion, so the placement of both the lowest and highest frequency eigenmodes will be key design parameters.

If the lowest frequency eigenmode of the quadruple suspension is designed to be around 0.5\,Hz, that means the soft mode frequency, $f_s$, will be about $-$0.1\,Hz. The required control bandwidth for this unstable mode would be less than 1\,Hz, enabling any sensing noise contribution from the soft mode to be suppressed by multiple orders of magnitude by 5\,Hz. Assuming that the highest frequency suspension eigenmode is around 1\,Hz, the hard mode frequency, $f_h$, would be around 1.5\,Hz, and the required control bandwidth for this loop would be around 2\,Hz. It is more difficult to achieve the required loop suppression for the hard loop without completely destabilizing the loop at the unity gain frequency. A suspension design that reduces the eigenmode frequencies further would be beneficial for the hard mode, but would risk increasing the bandwidth required for the soft mode control.

Furthermore, the aLIGO ASC is currently only sensing noise limited above 10\,Hz~\cite{ASC_MarieK}. Similar to the Michelson noise requirements, the displacement noise below 10\,Hz will need to be reduced to reach the \CE design goals at 5\,Hz. Again, improved local control from design aspects like the HoQI will reduce damping noise, and the proposed reductions of the seismic noise in Ref.~\cite{CE_LF} to the seismic noise will help further. However, the overall contribution of ASC noise will, in the end, depend heavily on the suspension design and resulting hard and soft mode frequencies. One way to improve ASC noise regardless of the resulting suspension design is to consider the sensing noise level. A factor of three reduction in sensing noise would reduce the injected ASC noise to the level of the MICH noise at 20\,Hz, as calculated in Section~\ref{sec:lsc}. This improvement would require increasing the power on the WFS to improve the shot-noise limit. Any reduction in the ASC sensing noise would help reduce requirements on the suspension design to achieve noise suppression at 5\,Hz.

The coupling of alignment noise in \CE must be studied more closely. The calculations in this paper assume the majority of the angular coupling from Eq.~\ref{eq:asc_coupling} would result from residual angular motion coupling to a static beam offset. Simulations will better reveal how \CE upgrades can improve the coupling of beam spot motion with residual angular motion, especially in regard to improved sensors and seismic isolation, as well as the significant suspension upgrades already proposed in works like Ref.~\cite{CE_LF}. Given the potential limitation to the gravitational-wave science from alignment control noise, it will be important to research its effect and various strategies for mitigation. For example, recent work projecting the current Virgo ASC design onto the Einstein Telescope design sensitivity curve demonstrates that significant improvements in ASC noise levels are required for XG detector goals~\cite{Maggiore:2024eun}.
At this time, improving the sensing noise could be the most effective method for reducing ASC noise, in concert with the already-proposed design choices for Cosmic Explorer. Since much of the coupling will also be related to the resulting shifts in the suspension modes due to radiation pressure, the effects of the alignment controls should be kept in mind during any suspension design work.

\subsection{Science impacts of low-frequency noise}

The technical noises highlighted in this section will have the most impact on detector sensitivity at low frequencies. Any loss in sensitivity at low frequencies compared to the expectation from fundamental noise sources will impact XG science. This section will detail a few impacts we foresee from any degradation in low-frequency sensitivity.

\begin{figure}[t]
  \centering
  \includegraphics[width=\textwidth]{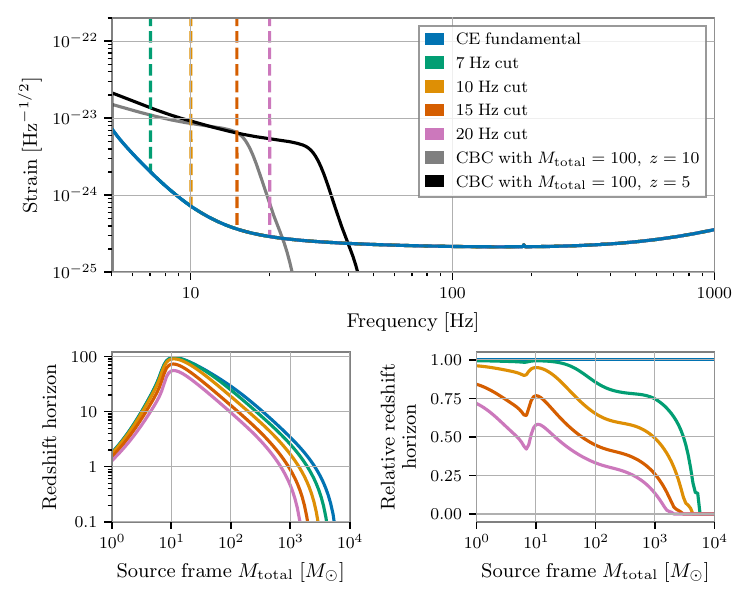}
  \caption{
  Top: The proposed~\cite{CE_noise_curve} \CE (CE) fundamental strain sensitivity with different low-frequency cut-offs overlaid. 
  Example compact binary coalescence (CBC) waveforms are also shown to demonstrate how different redshifts will affect the appearance of signals in the \CE detector.
  Signals from high redshifts will not be detectable by \CE as the signals will merge (in the detector frame) at frequencies lower than the detector's bandwidth.  \\
  Bottom left: The furthest distance a compact binary merger with a given mass can be observed with a signal-to-noise ratio of 8 for each low-frequency cut-off shown in the top panel. Source frame mass is quoted, as the measured mass at the detector will be redshifted. Masses in the compact binary are assumed to be equal. \\
  Bottom right: Relative redshift horizon for each low-frequency cut-off compared to the full-bandwidth \CE fundamental curve. Higher low-frequency cut-offs most impact the sensitivity to high-mass signals. 
  }
  \label{fig:hz_cuts_asd}
\end{figure}

To demonstrate the importance of low-frequency sensitivity, Figure~\ref{fig:hz_cuts_asd} shows how the total sensitivity of \CE is impacted by the lowest sensitivity considered. 
The upper panel shows the \CE design curve~\cite{CE_noise_curve} with the multiple low-frequency cut-offs we consider. 
The two lower panels show the redshift horizon\footnote{We calculate the redshift horizon~\cite{Chen:2017wpg} using the \texttt{inspiral-range} package~\cite{inspiral_range}, the IMRPhenomXPHM~\cite{Pratten:2020ceb} waveform, and the Planck18~\cite{Planck:2018vyg} cosmology.} of \CE with respect to the total source frame mass for each of these low-frequency cutoffs. 
The benefit of a higher detector bandwidth is most extreme for high-mass signals, but it also impacts the sensitivity of \CE to all signals. 
In the most extreme scenario, a low-frequency cut-off of 20\,Hz, there is over a 25\%  loss in sensitivity to all masses and a complete loss in sensitivity to masses above 2000\,$M_\odot$. 
For these high-mass systems, the higher frequency cut-off means that the merger frequency is lower than this cut-off in the detector frame due to redshift effects; 
this effect is demonstrated in the top panel of Figure~\ref{fig:hz_cuts_asd}.

In addition to the detectability of signals, we highlight how the low-frequency sensitivity of \CE will impact the estimation of source parameters from compact binary coalescences (CBCs).
A given detector's power spectral density, $S(f)$, is related to the signal-to-noise ratio (SNR) of a CBC signal (and hence the horizon distance of a detector) via~\cite{LIGOScientific:2019hgc}
\begin{equation}
  \text{SNR}^2 \propto \int \frac{f^{-7/3}}{S(f)} \text{d}f .
\end{equation}
However, the measurability of some CBC parameters is much more dominated by low frequencies than the SNR. 
For example, the measurability of chirp mass, $\Delta \mathcal{M}$, is given by~\cite{Damour:2012yf}
\begin{equation}
  \Delta \mathcal{M} \propto \int \frac{f^{-17/3}}{S(f)} \text{d}f .
\end{equation}
Due to the $f^{-17/3}$ frequency dependence, the loss in low-frequency sensitivity will have even more substantial impacts on parameter estimation than shown for the redshift horizon in previous sections. 

The loss in low-frequency sensitivity will also impact multi-messenger science. 
One of the exciting science goals that is unlocked with the increased bandwidth of XG detectors is the potential for warning of future signals well before their merger time. 
XG detector networks have the ability to confidently and accurately predict the sky location of impending merger over 10 hours in advance~\cite{Chan:2018csa,Li:2021mbo,Borhanian:2022czq}. However, as the time to coalescence, and hence the early warning timescale, is related to the lower frequency cutoff of the detector by~\cite{Sathyaprakash:1991mt}
\begin{equation}
  t_\text{warning} \propto f^{-8/3},
\end{equation}
the lowest frequency that XG detectors are sensitive to will strongly influence the potential early warning timescale. 

\begin{table*}[t]
    \centering
\begin{tabular}[\textwidth]{|l || c c |}
 \hline
 Detector configuration & $\text{SNR}_{1000}$ Horizon [Mpc] & Early warning [s] \\ 
 \hline\hline
 CE fundamental  &  113  &  1340  \\
 & & \\
CE, 7 Hz cut  &  113  &  931  \\
CE, 10 Hz cut  &  110  &  536  \\
CE, 15 Hz cut  &  100  &  214  \\
CE, 20 Hz cut  &  88  &  102  \\
 & & \\
 LIGO Livingston, O3  &  2.4  &  1.0  \\
 
 \hline
\end{tabular}
\caption{For the detector bandwidth cut-offs considered in Figure~\ref{fig:hz_cuts_asd}, we estimate the horizon distance at which a 1.4\,$M_\odot$ binary neutron star merger could be observed with a signal-to-noise ratio (SNR) of 1000 and the relative timescale before merger that a binary neutron star system located at a distance of 100\,Mpc could be well localized.
The early warning timescale is when each detector configuration could localize such a signal with the same precision as one second of early warning for LIGO Livingston in O3. 
In both cases the signal is assumed to be directly overhead the plane of the detector.
All \CE values have been rounded to the nearest integer. }
\label{tab:broadband_values}
\end{table*}
To demonstrate how different levels of broadband technical noise could impact
the SNR and the early warning timescale for binary neutron star (BNS) signals,
we estimate these quantities for a putative BNS signal with different low-frequency cut-offs. 
These results can be seen in
Table~\ref{tab:broadband_values}. We also include the same quantities for the
LIGO Livingston detector in O3 for reference. The horizon distance in this table
is the maximum distance at which a 1.4\,$M_\odot$ BNS signal could be observed at $\text{SNR}=1000$
in each detector configuration, and the early warning timescale is the duration
before merger that a similar system located at 100\,Mpc could be well-localized.
We define ``well-localized'' as similar in localization to 1\,s of early warning
for the LIGO Livingston detector in O3.

Other low-frequency science targets that will be heavily impacted include gravitational-wave displacement and spin memory effects~\cite{Grant:2022bla}, imprints of eccentricity on compact binary waveforms~\cite{Lenon:2021zac,Saini:2023wdk}, and the detection of persistent signals from rotating neutron stars (as will be discussed in Section~\ref{sec:lines}).
In general, the lowest frequencies accessible by XG detectors will be some of the most scientifically fruitful frequency regions. 
Maximizing the low-frequency sensitivity of XG detectors by addressing technical noise will be key to unlocking these science targets.

\section{Non-stationary and Transient Noise}\label{sec:trans_noise}

\begin{figure}[t]
  \centering
  \includegraphics[width=\textwidth]{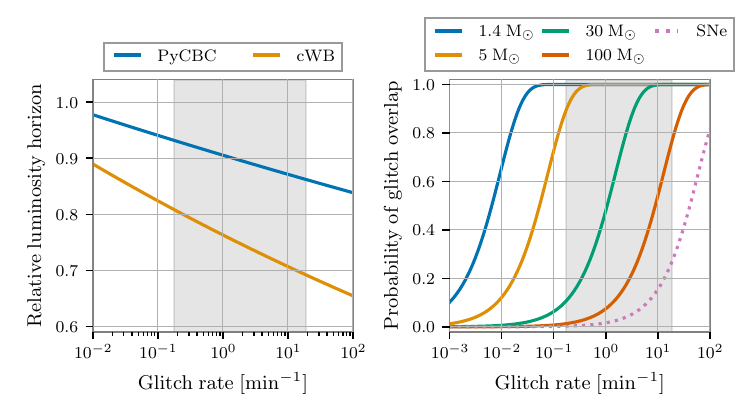}
  \caption{Estimates on how different glitch rates may impact future analyses of transient gravitational-wave signals. 
  The range of glitch rates achieved in Advanced LIGO, Advanced Virgo, and KAGRA detectors in O2 and O3~\cite{GWTC-3,KAGRA:2022twx} is indicated by the shaded gray region in both panels.
  Left: The relationship between glitch rate and the relative sensitivity (quoted as the relative luminosity distance of the horizon) for the PyCBC matched filter and cWB weakly modeled burst searches.  
  To meet the ideal sensitivity of XG detectors, the glitch rate must be reduced significantly compared to current detectors. 
  Right: 
  The probability of different low-redshift CBC signals overlapping a glitch with respect to the glitch rate.
  Each curve corresponds to a different mass value or a 1\,sec supernova (SNe) burst.
  For CBC signals, the signal is assumed to have a lower frequency of 5\,Hz. 
  To reduce the risk of glitch overlaps, the glitch rate must be reduced with respect to current detectors. 
  }
  \label{fig:glitch_impacts}
\end{figure}

So far, the technical noises we have discussed are both broadband and stationary.
Technical noise can also vary on the timescale of minutes to hours, as well as create short bursts of excess noise that last seconds or shorter~\cite{GW150914_detchar,Nuttall:2018xhi,LIGO_O3_detchar,Virgo_O3_detchar,KAGRA_detchar}. 
The long-timescale variations are referred to as non-stationary noise, while short bursts are referred to as transient noise or ``glitches.''
Non-stationary and transient noise sources are harder to quantify than stationary ones but could also significantly impact the possible science from a detector like Cosmic Explorer. 

A wide variety of sources of non-stationary or transient noise have been identified in the current generation of gravitational-wave detectors; 
as an example of these sources, two of the most significant sources of non-stationary or transient noise in aLIGO include scattered light and environmental noise~\cite{Accadia:2010zzb,Soni:2020rbu,AdvLIGO:2021oxw,Virgo:2022ypn,KAGRA_detchar}. 
Scattered light, as the name suggests, is from light that is scattered off of the main laser beam path due to imperfections in the optics~\cite{Accadia:2010zzb,Soni:2020rbu,Soni:2023kqq}.
This light then reflects off a moving surface and rejoins the main beam path with a different phase, interfering with light in the main beam path and creating noise in the gravitational-wave strain data. 
As scattered light is only impactful if it reflects off of a moving surface, sources of excess motion from the environment (generally in the form of seismic motion) can induce scattered light that creates transient noise for multiple hours~\cite{Accadia:2010zzb,Soni:2020rbu,Soni:2023kqq}.
In addition to variable seismic motion, other sources of non-stationary environmental noises include acoustic impacts from thunder~\cite{LIGO_O3_detchar}, magnetic coupling effects from lightning~\cite{GW150914_detchar,Washimi:2021ogz,Virgo:2022ypn,Janssens:2022tdj}, radio frequency interference~\cite{GW150914_detchar,LIGO_O3_detchar}, and even induced vibrations from local wildlife~\cite{2023ApPhL.122r4101B}.
The coupling mechanism between environmental noise and the gravitational-wave strain data can be difficult to identify due to the large size and complexity of gravitational-wave interferometers; 
for this reason, observatories utilize large sensor arrays and careful measurements of localized coupling functions to identify any problematic sources of environmental noise~\cite{AdvLIGO:2021oxw,Virgo:2022ypn,KAGRA_detchar}.

At present, the most concerning type of technical noise for analyses of short-duration gravitational-wave signals is transient noise.
Transient noise sources impact current these astrophysical analyses in two key ways: 
biasing analyses of detected gravitational-wave events or preventing their identification altogether. 
Both of these challenges have limited current analyses and have led to extensive work to mitigate transient noise sources in current detectors~\cite{LIGO_O3_detchar, Virgo_O3_detchar, KAGRA_detchar}.
Guided by the impacts of glitches on recent analyses, we estimate how these technical sources of noise may impact XG analyses.

The presence of transient noise itself makes it more challenging to identify transient gravitational-wave signals, as it is more difficult to differentiate astrophysical signals from the background of instrumental artifacts\cite{TheLIGOScientific:2017lwt,LIGO_O3_detchar}. 
To address this, modern gravitational-wave searches employ ``signal consistency tests'' that improve discrimination between signals and noise~\cite{Allen:2005fk,Nitz:2017lco,Mozzon:2020gwa,Messick:2016aqy,Zackay:2019kkv}. 
Despite this, glitches still significantly impact the sensitivity of searches. 
We use a suite of injections performed as part of the GWTC-3~\cite{GWTC-3} analyses to estimate the sensitivity of search pipelines in O3 to evaluate how the rate of glitches can impact the sensitivity of the search. 
We focus on two pipelines: the PyCBC matched-filter search~\cite{Usman:2015kfa} and the Coherent WaveBurst (cWB) weakly modeled search pipeline~\cite{Klimenko:2008fu}.
These two searches were chosen as they analyze data in approximately two-week chunks, allowing us to evaluate how different glitch rates impact their sensitivity, and to include representative matched filter and burst searches. 
Our results are shown in Figure~\ref{fig:glitch_impacts}.
The full details of how we estimate the impact of glitches on sensitivity are included in \ref{app:glitch_rate}.

We find that the glitch rate in XG detectors has the potential to be one of the limiting sources of noise, with comparable impacts to broadband technical noise. 
For current analyses and the O3 glitch rate, we find that the presence of glitches reduces the sensitivity of the searches to systems above 100\,$M_\odot$ in the detector frame (compared to the ideal case) by $\sim$10\% for PyCBC and by $\sim$23\% for cWB.
There are significant differences across the mass range probed by these searches; 
at low masses (e.g., BNS systems), there is almost no impact due to glitches, while at even higher masses, the impact increases~\cite{O2-DQ-paper,Godwin:2020weu}. 
This is likely due to the difficulty in differentiating a short astrophysical signal from a short instrumental artifact. 
This mass-dependence is supported by our own analysis; 
we find no correlation between the glitch rate and the sensitivity of search pipelines for BNS systems, compared to a strong relationship for high-mass binary black hole (BBH) systems. 
The effect of glitches on the sensitivity of searches for even higher mass systems is likely even stronger than for 100\,$M_\odot$, but we were not able to estimate this correlation due to a lack of data in this injection set in this mass parameter space.
We also note that there is a stronger correlation between the glitch rate and the loss in sensitivity for cWB than for PyCBC. 
Hence, there is more to gain for burst searches by reducing the glitch rate in future detectors. 

The network of detectors that are operating in unison during the XG era will greatly influence how glitches impact XG analyses. 
In the case that only a single detector is operating, one can no longer use coincidence between multiple detectors to identify astrophysical signals in a data stream full of non-coincident glitches. 
Therefore, any network that contains multiple detectors will be much more sensitive to signals that could be mimicked by glitches~\cite{Beauville:2005au}. 
The sensitivity estimates in Figure~\ref{fig:glitch_impacts} are based on O3 data where up to three detectors were operating. 
The sensitivity impact would likely be much worse than estimated here for the 1-detector case. 
One potential exception would be a detector consisting of multiple co-located interferometers with different responses to astrophysical signals, such as the Einstein Telescope. 
In this case, it may be possible to combine the individual interferometer data streams to effectively separate glitches from astrophysical signals~\cite{Goncharov:2022dgl}.

Although all gravitational-wave signals to date have been identified with at least one matched filter search~\cite{GWTC-3}, the impacts on burst searches, such as cWB, are particularly important for XG analyses due to the effects of redshifting on cosmological sources. 
At the horizon distances probed by XG detectors, even the lowest mass black hole systems would appear over 100 $M_\odot$ in the detector frame. 
At these masses, current burst searches begin to meet or exceed the sensitivity of matched filter searches~\cite{O2_IMBH}. 
Hence, for all but the lightest systems, it is possible that the highest redshifts will be probed by burst searches. 
To ensure that cosmological, transient sources of gravitational waves are detectable with XG detectors, 
it is key that the glitch rate is minimized as much as possible and that there is additional investigation of how to maximize both matched filter and burst searches for high mass CBC sources. 

These estimates of glitch impacts on search sensitivity can only be extrapolated to the XG detector era assuming that the process of detecting gravitational-wave signals is similar to the current detection procedure. 
Improvements to the search pipelines, 
a much higher event rate, 
and new methods of identifying signals will all impact the level of impact glitches will have on searches for gravitational waves in the XG era. 
However, the limitations that glitches currently cause for search sensitivity highlights that a reduced glitch rate as well as new methods to mitigate their impact will be beneficial to XG science.

The second main way that glitches can impact astrophysical analyses is by direct overlap with real signals. 
As analyses of gravitational-wave signals typically assume that noise in detectors is stationary and Gaussian~\cite{LIGOScientific:2019hgc}, the presence of glitches severely violates these assumptions and can create biases~\cite{Powell:2018csz,Kwok:2021zny,Mozzon:2021wam,Macas:2022afm,Hourihane:2022doe,Ghonge:2023ksb,Davis:2022ird}.
Since a subset of signals will overlap glitches regardless of the glitch rate, precise glitch modeling and subtraction is essential to probe interesting astrophysical events.
While it is generally straightforward to identify the presence of glitching, for low-SNR glitches, it is challenging to differentiate Gaussian noise fluctuations from these artifacts, potentially increasing the risk of biases. 
This can lead to ambiguity in whether low-SNR glitching leads to biases in analyses~\cite{Payne:2022,Macas:2023wiw}.
Increasing the difficulty of this problem is the large number of different glitches and the constant evolution in the types of glitches in the detector data. 
For this reason, it is beneficial to have glitch subtraction tools that are designed for specific glitch classes~\cite{Torres-Forne:2020eax,Merritt:2021xwh,Udall:2022vkv,Tolley:2023umc,Bondarescu:2023jcx}, specific scenarios (such as when a different sensor witnesses the source of the noise)~\cite{Davis:2019,Davis:2022ird,Macas:2023wiw}, as well as generic methods~\cite{Cornish:2014kda,Pankow:2018qpo,Cornish:2020dwh,Hourihane:2022doe, Zackay:2019kkv, Ashton:2022ztk}.

To assess the impact of this problem for XG detectors, we estimate the probability that a glitch will overlap a signal for an XG detector based on different glitch rates. 
We note that a glitch overlapping a signal does not inherently mean that analyses of this signal will be biased; 
this estimate is, therefore, an upper limit on the level of biases that may be created. 
To estimate the probability of a glitch overlapping a signal, we assume that glitches are a Poisson process and that all gravitational-wave signals have a lower frequency cutoff of 5\,Hz.
The estimates for this calculation for various glitch rates and signal parameters are shown in the right panel of Figure~\ref{fig:glitch_impacts}.
For current glitch rates, essentially all BNS signals will overlap a glitch, while overlaps for larger BBH systems are rare but non-trivial. 
If glitch rates increased by an order of magnitude, glitch overlaps would be non-trivial for events with the highest masses and short burst signals, such as SNe.
Preventing overlaps with BNS signals would require more than an order of magnitude reduction in the glitch rate. 
Current estimates for the rate of BBH signals in \CE are comparable to the current rate of glitches. 
Hence, the risk of glitch overlaps should be considered comparable or more concerning than the risk of overlaps with astrophysical signals, even when the highest event rates are assumed. 

Numerous scientific objectives with XG detectors will require low glitch rates to be achievable. 
Detectors with low glitch rates will be vital to identify GW signals from the furthest redshift accessible with \CE, such as population III stellar remnants~\cite{Sesana:2009wg} and the earliest mergers of primordial black holes (PBH)~\cite{Carr:2020xqk}. 
A low glitch rate will also minimize the risk of missing exciting signals or biased analyses. 
This is particularly vital for short, rare signals such as supernovae (SNe) that are expected to be detectable with XG detectors~\cite{Srivastava:2019fcb,Vartanyan:2023sxm}.

It should be noted that for glitch rates observed in current-generation detectors, the chance of a signal directly overlapping a glitch is comparable to or higher than the chance of two signals overlapping. 
Depending on the astrophysical assumptions, one would expect a $\mathcal{O}$(10\%) of binary black hole signals to overlap~\cite{Pizzati:2021apa,Himemoto:2021ukb,Hu:2022bji}.
As seen in Figure~\ref{fig:glitch_impacts}, the chance of a glitch overlapping is similar to, or higher than, this expectation. 
Glitches are also broadband in nature and unmodeled, which may create additional challenges compared to the overlap of two signals that are well-modeled.

Conversely, broadband non-stationarity of the detector noise sensitivity is likely to not be a major impediment to the core science cases of XG detectors. 
Compact binary signals will be in band much longer for XG detectors than current detectors, meaning that it will be essential that non-stationarity is accounted for when estimating the properties of these signals. 
However, methods have already been developed to address this challenge~\cite{Mozzon:2021wam,Kumar:2022tto}, as current-generation detectors already exhibit high levels of non-stationarity. 
Furthermore, such data quality issues are routinely addressed in other data analysis techniques, such as searches for gravitational-wave signals~\cite{gstlal-methods,Zackay:2019kkv,Mozzon:2020gwa}.

In addition to impacting data quality, non-stationary and transient noise can prevent data from being recorded at all. 
Particular large disturbances or detector instabilities can lead to the detectors losing resonance of the laser inside the optical cavities --- often referred to as a ``lock loss.''
After a lock loss, the operation of the detector must be restored, leading to sometimes significant downtime when no observations are possible. 
There have been lengthy investigations into the causes of lock losses~\cite{Biswas:2019wmx,o3LIGOinstrumentation}
and development of potential mitigation strategies~\cite{Coughlin:2016lny, Biscans:2017yce,LIGO:2020pzq}.
If there are a large number of lock losses (which would lead to a low duty cycle), there is an increased risk of missing rare transient events such as SNe, nearby electromagnetically bright signals~\cite{GW170817}, or signals that are observable in multiple gravitational-wave bands~\cite{Colpi:2016fup,Gerosa:2019dbe,Toubiana:2022vpp}.
A reduced duty cycle will also reduce the total detection rate and lead to reduced precision in astrophysical analyses due to the reduced number of detectors that observe any given event. 
The reduced number of detectors will be particularly impactful for sky localization of events, as the time of arrival of the signal in each detector is one of the main sources of localization information~\cite{Fairhurst:2010is}. 

Even when a detector is observing, it is possible for non-stationary and transient noise to be so detrimental to the data quality that the data is no longer fit for astrophysical analysis.
For example, the rate of glitches may be so high that it would be impossible to identify an astrophysical transient in the data or the detector noise level is not able to be accurately measured due to the high level of non-stationarity. 
In recent observing runs, it has been typical for a few percent of data to not be analyzed even when the detector was observing~\cite{GW150914_detchar,LIGO_O3_detchar}~\footnote{This includes times when the detector was misconfigured in addition to times when the data quality was poor~\cite{T2100150,T2100045}.}.
It is also possible for low-SNR data quality effects to render data unusable for a subset of astrophysical analyses.
As an example of how subtle effects can have a significant impact, five days of data were not analyzed in O3 due to the presence of a camera creating glitches from the vibration of the shutter at a regular cadence~\cite{Goetz:2023aaa,Goetz:2021aaa}.
These glitches, combined with additional data processing, created such significant combs in the data near lines that it was not beneficial for many persistent gravitational-wave searches to analyze this data.

\section{Narrow Noise Features}\label{sec:lines}

The presence of persistent narrow noise features in XG detectors will be one of the main limitations to the discovery of new sources of persistent gravitational waves and hamper the analysis of known signals. 
Often referred to as ``lines,'' these near-sinusoidal instrumental artifacts are present in all current gravitational-wave detectors at frequencies across the entire band of interest. 
There are a wide variety of different sources of lines, all of which have the potential to hamper astrophysical analyses~\cite{LSC:2018vzm,LIGOScientific:2020qhb,LIGO_O3_detchar,LIGOScientific:2017csd,Tenorio:2021wmz}.

\begin{figure}[t]
  \centering
  \includegraphics[width=\textwidth]{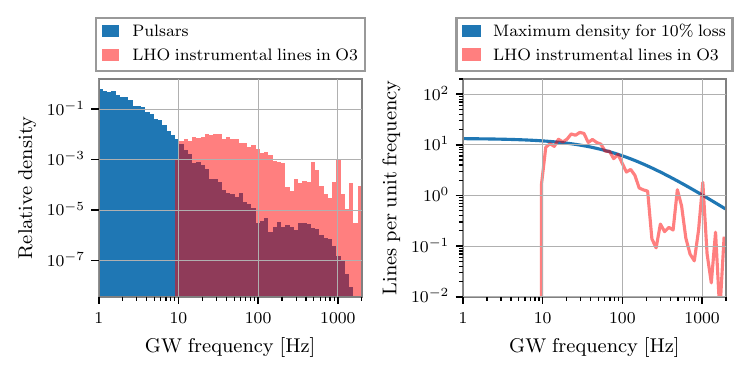}
  \caption{A comparison of the density of gravitational-wave signals from pulsars and instrumental lines in the LIGO Hanford (LHO) detector during O3. 
  Left: the relative density of signals from pulsars and line artifacts per unit frequency.
  No lines were recorded below 10~\,Hz as data from this frequency region is not considered usable. 
  Note that the density of both pulsars and lines grows at lower frequencies.
  Right: a comparison in the number of lines per unit frequency and the maximum number of lines allowable per unit frequency for 10\% of pulsars to be undetectable due to lines. 
  Both the physical width of lines and the Doppler broadening of the signal are considered. 
  In O3, over 10\% of pulsars were already not observable at some frequencies.
  Pulsar frequencies are derived using data from the Australia Telescope National Facility Pulsar Catalog~\cite{Manchester:2004bp,ATNF_catalog}.
  LHO instrumental lines based on data from the Gravitational Wave Open Science Center~\cite{GWOSC_O3,O3_lines}.
  }
  \label{fig:line_impacts}
\end{figure}

While these types of instrumental artifacts may not impact analyses of transient
gravitational-wave sources, searches for persistent gravitational-wave signals
can be severely impaired by data containing narrow-band
features~\cite{LIGOScientific:2020qhb,LSC:2018vzm,Tenorio:2021wmz}. The most well-known source of persistent gravitational waves is rapidly spinning neutron stars~\cite{Riles:2022wwz}. As shown in the left panel of Figure~\ref{fig:line_impacts}, the density per unit frequency of known spinning neutron stars (detected as pulsars) is inversely correlated with the gravitational wave frequency. 
Unfortunately, the same is true of instrumental lines. The frequency width impacted by instrumental lines comprises two effects: the physical width of the instrumental line and the Doppler broadening~\cite{Wette:2023dom} of the astrophysical signal. 
As it is common to veto any candidate where the frequency of the signal overlaps an instrumental line after Doppler broadening is concerned, we can consider the entire frequency region lost~\cite{Tenorio:2021wmz}. 

At low frequencies, the physical width of the line dominates the bandwidth that is impacted. 
This can be seen in the right panel of Figure~\ref{fig:line_impacts}; as $f\rightarrow 0$, the maximum density of lines that is allowable for a 10\% loss in frequency space plateaus based on the assumed width of lines. 
At high frequencies, the effect of Doppler broadening dominates, leading to stricter requirements for a fixed loss of bandwidth per frequency. 
Even though the density of lines does not grow at low frequencies as quickly as the density of pulsars, the fixed maximum density of lines means that there is a risk of significant loss in discovery potential unless the density of lines is reduced below current levels. The presence of narrow-band features is already a problem for current gravitational-wave detectors, and the lower noise floor of XG detectors will exacerbate this issue in the absence of mitigation.

Sources of lines can be
roughly divided into three categories: purposefully injected lines, such as
calibration lines or dither lines; well-understood lines related to the detector design, such as
mechanical resonances; and spurious lines resulting from technical and
environmental noise couplings.
Many lines only impact a narrow
frequency band, but some lines can be found in so-called ``combs,'' where groups
of lines are present at evenly spaced frequencies and have a shared physical source~\cite{LSC:2018vzm,Virgo_O3_detchar}.
Spurious lines without evidence for an instrumental origin are particularly problematic; 
such lines are challenging to monitor and can be confused with an astrophysical signal more easily than those with a well-understood source.
Ongoing monitoring and investigation of lines will be essential to address spurious lines that will appear during the operation of XG detectors. 

Any lines that do not have explicit evidence of instrumental origin will be especially problematic for searches of persistent signals
as lines with such evidence can be safely discarded as candidates for an astrophysical signal. 
Discarding candidates related to unknown sources may unintentionally veto a true signal, resulting in a trade off between polluting a search with instrumental lines that do not have evidence of their origin versus risking missing an astrophysical signal. 

In addition to the challenges presented by individual lines, mixing (or ``modulation'') of different instrumental lines has been observed in current detectors. 
When two lines with frequencies $f_1$ and $f_2$ mix, they can create additional lines at frequencies $|f_1+f_2|$ and $|f_1-f_2|$.
Injected lines like dithers, mechanical
resonances, calibration
lines and spurious lines have been observed to modulate with each other and spoil additional frequency
bands~\cite{o3LIGOinstrumentation, LSC:2018vzm, LIGO_O3_detchar}.
Mixing can also occur between narrow lines in the gravitational-wave frequency band and low-frequency sources, such as seismic noise~\cite{LIGO_O3_detchar}.
Even lines that are well above the gravitational-wave frequency band, such as dither lines injected at around 2 and 4\,kHz, have been known to mix with calibration lines or mechanical resonances to create artifacts at frequencies of astrophysical interest~\cite{o3LIGOinstrumentation, LSC:2018vzm, LIGO_O3_detchar}.
Line mixing is particularly problematic as it is can arise unexpectedly, allowing well understood lines to create additional spurious lines. 
As mentioned, mixing is also one way that injected lines or mechanical resonances that are intentionally placed outside a specific frequency band can spoil frequencies of interest.

Noise from purposefully injected lines can be thought of as a technical noise similar to broadband controls noise in that they are required for the successful operation of the detector. In O3, both LIGO detectors used Alignment Dither System (ADS) lines~\cite{o3LIGOinstrumentation}. ADS lines maintain the beam spot position on the test masses, a requirement for maintaining stable alignment controls. A dither line from each test mass is injected into the DARM signal and then demodulated to obtain an angle-to-length coupling signal that is minimized when the beam position is co-aligned with the optic rotation point~\cite{o3LIGOinstrumentation}. Both detectors injected these lines below 20\,Hz, which has the potential to impact low-frequency persistent gravitational-wave searches.  Nothing in the initial \CE design~\cite{CE_horizon} nominally requires dither schemes over other alignment or locking schemes, but they may become necessary for successful detector operation.
If dither schemes are required for XG detectors, careful consideration of the location and amplitude of dither lines should be taken to avoid contaminating searches as much as possible and careful monitoring for modulation effects should be planned.

Calibration lines are another set of purposefully-injected lines. These
lines are used to track slowly occurring changes in the response of the
detector. While these time-varying changes are small enough that they do not impact the operation of
the detector, they may cause systematic biases in the
calibrated gravitational-wave strain data if
uncompensated~\cite{Tuyenbayev_2016,Viets_2018}. The presence of uncompensated
systematic biases in the calibrated strain data would negatively impact
astrophysical parameter estimation of gravitational-wave
sources~\cite{Essick_2022,Cahillane:2017vkb}. Calibration lines are loud, known
sinusoidal excitations whose detector response amplitudes and phases are tracked
over time and used to correct the static calibration model to account for time
dependence. The resulting correction factors are called ``time-dependent
correction factors'' (TDCFs)~\cite{Tuyenbayev_2016,Viets_2018}. Both LIGO detectors use
TDCFs calculated from calibration line injections for calibration purposes: to
track the actuation and sensing components of the detector response function.
While their frequencies may differ between LIGO Hanford and LIGO Livingston, the
lines serve the same purpose at both facilities~\cite{Sun_2020,Viets_2018}. The
calibration lines used to track the detector's actuation function are all placed
between 10 and 20\,Hz, where the actuators in the lowest three stages in the
quadruple pendulum are comparable in magnitude. The detector's optical gain and
coupled-cavity pole frequency are tracked using a line near the center of aLIGO's
most sensitive frequency band; this is where their effects are most
significant~\cite{Viets_2018}. Also tracked are additional sensing function
parameters to characterize the opto-mechanical effect of operating a detuned
signal extraction cavity ($\sim$8\,Hz)~\cite{Cahillane:2017vkb,Sun_2020}. Line
height is determined by a desired SNR for calibration
precision, the limitations of the injection system, and consideration for
mitigating often unpredictable contaminations, such as those caused by linear
and non-linear frequency mixing. 
While calibration lines will be a
necessary part of detector operation, their placement, quantity, and amplitude
will require consideration for the next generation of detectors due to the
promise of detection of persistent gravitational waves. Calibration is
discussed further in Section~\ref{sec:cal}.

Mechanical resonances of optic suspensions also appear as lines in DARM. In Advanced LIGO, the suspension chains are designed to have very low loss, so any mechanical resonance of that suspension will have a high quality factor (Q)~\cite{o3LIGOinstrumentation}. The same is expected for \CE~\cite{CE_LF, CE_horizon}. For example, resonances of the fused silica suspension fibers used to suspend the test masses and beamsplitter (often referred to as ``violin modes'') appear in the DARM spectrum as well as their harmonics~\cite{o3LIGOinstrumentation}. 
In the LIGO detectors, these suspension resonances can create severe line contamination when they are rung up by seismic activity~\cite{alog:violin_lines}.
Other mechanical resonances, such as the test masses' vertical and roll mode resonances, can appear in DARM. While the most significant mechanical resonance lines in aLIGO have resulted from the suspensions of the test masses, in principle, mechanical resonances from auxiliary optics could also contaminate the spectrum. With the increased sensitivity of \CE, these mechanical resonances can become prominent and contaminate other portions of the spectrum. Unique to \CE design is the potential to also see the presence of high-Q test mass acoustic modes within the detector bandwidth due to the larger size of the test masses relative to aLIGO.

Mechanical resonances can be passively or actively damped~\cite{robertson_BRD}, but that is usually done to mitigate the effect these lines have on locking or control loops, as these resonances can become excited in amplitude and saturate interferometric sensors~\cite{o3LIGOinstrumentation}. Mechanical damping of these resonances lowers the quality factor of the resonance, which could broaden the peak in the DARM spectrum. Acoustic modes are also damped passively to avoid parametric instabilities that cause locklosses~\cite{o3LIGOinstrumentation, Biscans_AMD}. Like mechanical damping, acoustic mode damping also lowers the quality factor of the mode~\cite{o3LIGOinstrumentation}. Depending on the location of the resonance and the relative broadening of the peak, it may not be favorable to persistent gravitational-wave searches to add these types of damping mechanisms. However, damping may be required as a part of detector operation. Therefore, if possible, consideration of the locations of mechanical resonances should occur during the design stage of a new detector like Cosmic Explorer, as no mitigation technique applied during detector operation will be able to eliminate them entirely. Better witnesses of this mechanical motion could allow subtraction of the lines, but that may not account for line modulation unless non-linear effects, such as those introduced from digital-to-analog conversion, are considered~\cite{Driggers:2018gii,Davis:2019,Vajente:2019ycy,Viets:2021aaa,Yu:2021swq}.

Finally, current generation detectors have also witnessed contamination from ``spurious'' lines that result from unintended noise couplings~\cite{LSC:2018vzm,LIGO_O3_detchar,Virgo_O3_detchar}. 
The sources of these lines are often traced to weak coupling between electronics operating at the observatory or another periodic source of noise.
For example, aLIGO has experienced significant impacts from synchronized 1\,Hz blinking of LED lights on various electronics, magnetic fields produced by vacuum sensors, and influences from ethernet cable adaptors~\cite{LSC:2018vzm,LIGO_O3_detchar}.
Due to the low amplitude of these instrumental artifacts, identifying them requires extensive investigations over multiple months. 
It also requires long time periods to evaluate if any technical changes have mitigated the source of noise~\cite{LSC:2018vzm}. 
To ensure that these types of lines have minimal impact on analyses, constant monitoring of the data is required, along with careful record keeping; 
since the effects of instrumental changes may not be noticed for months, it is essential that it is possible to precisely correlate (sometimes down to the minute) the time of appearance of an artifact with on-site activities. Similar to mechanical resonances, much of the mitigation of these lines is ideally done at the design stage, as better isolation (electronic, magnetic, and mechanical) can prevent these lines from ever appearing.
However, these examples demonstrate that not everything can be reasonably anticipated in the design of an observatory; 
for this reason, constant monitoring of line sources and coordination between data analysts and on-site investigators will be required throughout the era of XG to catalog or mitigate new lines as they appear.  

Due to the significant discovery risk that narrow-band features present to searches for persistent gravitational waves, additional research in understanding the causes of instrumental lines and how to address them in data analysis pipelines is prudent. 
In particular, methods of how to mitigate spurious lines through instrumental design are not well represented in the literature and will be important to understand more clearly in the near future. 
Furthermore, planning for XG observatories should explicitly include support for long-term monitoring and mitigation of narrow noise features.

\section{Noise correlated across detector networks}\label{sec:corr_noise}

In addition to noise that is present only in a single interferometer, it is worth considering instrumental noise that may be correlated across the gravitational-wave detector network. 
Such noise is particularly concerning for any astrophysical analysis, as the presence of correlated noise violates the assumed independence between detector data streams and may mimic an astrophysical signal. 
For this reason, care has been taken to evaluate the current level of correlated noise and understand its impact on analyses~\cite{GW150914_detchar,Thrane:2014yza,Coughlin:2016vor,Coughlin:2018tjc,Meyers:2020qrb,Janssens:2022tdj}.
The analyses that are most impacted by correlated noise are searches for stochastic signals that rely on cross-correlating multiple detector data streams to identify a gravitational wave background~\cite{Renzini:2022alw}.
In this section, we discuss the current understanding of correlated noise in ground-based detectors and the implications for XG detectors and stochastic analyses. 

Current estimates of the population of gravitational-wave signals from CBCs suggest that the detection of an astrophysical stochastic background from gravitational waves is possible with the current generation of facilities~\cite{KAGRA:2021kbb,GWTC3_pops}.
It is, therefore, unlikely that any technical noises would prevent the identification of a similar background with much more sensitive XG facilities. 
However, XG facilities also have the potential to identify stochastic backgrounds from cosmological sources or from novel emission mechanisms of gravitational waves~\cite{Renzini:2022alw,CE_horizon,ET_science}. 
Stochastic gravitational-wave sources represent one of the most interesting potential ways that unexpected discoveries could be made with XG facilities.
Any technical noise that would impact the identification of a stochastic signal could prevent such discoveries.

At the time of writing, there have already been multiple works devoted to investigating the level of correlated noise that is expected in XG facilities from magnetic sources, such as Schumann resonances and lightning strikes in the context of Einstein telescope~\cite{Janssens:2021cta}.
In this work, the authors conclude that increased magnetic isolation by two to four orders of magnitude (compared to current detectors) is required to prevent magnetic noise from impacting stochastic analyses. 
If this is not possible, it may be possible to subtract magnetic noise using e.g., Wiener filtering~\cite{Thrane:2014yza,Coughlin:2016vor,Coughlin:2018tjc}.
There is also ongoing work to account for unsuppressed magnetic noise in stochastic analyses that may further reduce the risk of magnetic impacts on stochastic analyses~\cite{Meyers:2020qrb}.
Any analysis that relies on co-located detectors may also have to contend with other sources of correlated noise~\cite{Janssens:2022xmo}, although it may be possible to address this in analyses by combining detector data to create a null stream that is insensitive to gravitational waves~\cite{Wong:2021eun,Janssens:2022cty,Goncharov:2022dgl,Cireddu:2023ssf}.

With the increased sensitivity of XG detectors, there is an increased risk of correlated transient noise in addition to correlated stochastic noise. 
One potential source of such noise is lightning strikes; such strikes have already been observed to create instrumental artifacts in the KAGRA~\cite{Washimi:2021ogz} and Virgo~\cite{Virgo:2022ypn} detectors. 
One such study has been done into the risk of correlated instrumental artifacts from lightning correlations~\cite{Janssens:2022tdj}. 
This would be particularly important for detectors that are located on the same continent or within range of the same lightning sources. 
However, both the magnetic effects of lightning and the vibrational impacts of thunder are well witnessed by sensors located at each observatory~\cite{Washimi:2021ogz,Virgo:2022ypn,AdvLIGO:2021oxw}, making it unlikely that related instrumental artifacts would be mistaken for astrophysical signals.

\section{Calibration}\label{sec:cal}
As with Advanced LIGO, Advanced Virgo, and KAGRA, future gravitational-wave
detectors will need to convert raw detector output into estimates of
gravitational-wave strain. The LIGO detectors require closed-loop feedback
control of the differential arm length degree of freedom for stable
operation~\cite{o3LIGOinstrumentation}. We will refer to this closed-loop system, which is implemented in
the LIGO detectors as a feedback servo, as the ``DARM loop'' in this
paper~\cite{Sun_2020}. At its core, the DARM loop consists of an opto-mechanical
``sensing'' component and a feedback actuation component. Calibration of the
strain incident on the detector relies directly on the accuracy with which the
sensing and actuation DARM loop components are modeled and accounted for. In
practice, signal processing electronics and the force-to-length dynamics of each
of the masses in the actuation chain must be properly characterized and
compensated for.

The accuracy and precision with which the reconstructed strain data is
calibrated is important for the detection of gravitational waves and critical
for the estimation of astrophysical parameters~\cite{Lindblom:2009un,Viets_2018,Abbott_2017}. The deviation of the
reconstructed strain data from the true strain incident on the detector is known
as the systematic error in the calibration~\cite{Sun_2020}. Contributors to this error
include known systematic issues with imperfect compensation, unknown systematic
errors, as well as fundamental error in the calibration fiducials. In LIGO, the
calibration fiducial error is $<1\%$, but this requires constant and dedicated
attention to maintain~\cite{galaxies10020042}. During O3, the highest levels of systematic error and
uncertainty reported in the calibration of the Advanced LIGO and Advanced Virgo
gravitational-wave detectors stood at $\sim 10$\% in magnitude and $\sim 10^{\circ}$ in
phase in parts of the sensitive frequency band 20\,Hz to 2\,kHz~\cite{sun2021characterization,Virgo:2021kfv}. Such a
level of systematic error has remained tolerable because decreasing the combined
calibration and waveform-modeling errors would be made irrelevant by noise in
the detector~\cite{Lindblom:2009un,Vitale:2011wu,Hall_2019,Payne_2020,Vitale_2021,Huang:2022rdg,Wade_2023}.

The vastly improved sensitivity that \CE{}--like detectors promise to deliver is
expected to be accompanied by detections with SNRs in the thousands~\cite{CE_horizon}. For
comparison, the loudest binary black hole detection in GWTC-3 is GW200129 with
SNR $\approx 26$~\cite{GWTC-3}. \CE{}'s improved sensitivity, combined with the improvements in
theoretical waveform accuracy that it will drive, will prompt requirements in
calibration error that are much more demanding than what is in place with
today's detectors. The combined effects of calibration errors, high SNR signals,
and the accuracy and precision of parameter estimation analyses are actively
being explored~\cite{Read_2023,Hall_2019,Payne_2020,Vitale_2021}. The exact calibration requirements of the \CE{}
detectors are not yet known. Nonetheless, it is expected that accuracies
significantly below $1\%$ will be necessary to avoid limiting scientific
output~\cite{CE_horizon}.

If achieved, calibration uncertainty of less than 1\% in \CE{}--like detectors
will bolster a multitude of astrophysical science objectives including tests of
general relativity and estimation of the Hubble constant with reduced
uncertainty~\cite{Chen_2021,Hall_2019}. In addition, there is a strong scientific motivation to
maintain and extend these levels of calibration accuracy in a frequency range
spanning from below 20\,Hz to a few kilohertz. Improving the calibration
accuracy at low frequencies will enable precision tests of General Relativity
via the analyses of black hole ringdown signals and support measurements of
compact binary inspirals at cosmological distances~\cite{Srivastava_2022,Schutz_2018}. It was recently
estimated that the error due to calibration and waveform modeling associated
with \CE{} will need to be close to $0.1\%$ at 20\,Hz in order for calibration
to remain subdominant to stochastic noise in the presence of a BNS signal at
100\,Mpc~(see Figure 5 in Ref.~\cite{Read_2023} and related discussion). Such a system is
expected to have an SNR $\simeq 1158$ above 10\,Hz with a single \CE{} detector~\cite{Read_2023}.
Taking the same accuracy requirements up to a few kilohertz will allow for
tighter constraints to be placed on the nuclear equation of state of neutron
stars using measurements of post-merger gravitational waves~\cite{Hernandez_Vivanco_2019, Srivastava_2022,Chatziioannou_2020}.
Gravitational-wave observations above $\sim$2\,kHz would permit the exciting
possibility of detecting new gravitational-wave sources such as supernovae or
isolated pulsars~\cite{Srivastava_2022}.

Today's calibration techniques will need to be extended significantly to meet
the requirements of \CE{} and other XG gravitational-wave detectors.
For example, the calibration method employed currently for the LIGO detectors is
predominantly limited by: 1) variations in the optical plant not currently
accounted for by the optical plant model and 2) difficulty in measuring the
detector response above 1 kHz using the relatively weak current-generation
length fiducial actuators. These two limitations generally result in an
undesirable increase in calibration uncertainty at low and high frequency of the
detection band~\cite{Sun_2020,sun2021characterization}. Each of these components will rely on future research and
development to usher them into the era of \CE{} and to meet the strict
requirements on calibration error and uncertainty that will be required~\cite{CE_horizon}. In
addition to the R\&D needed to build upon the calibration methods currently in
use, future work will be needed to evaluate the scalability of today's
calibration implementation methods with the long term sustainability and
maintainability of the continuous operation next generation detectors in mind.
The remainder of this section briefly discusses these components and the
obstacles that will need to be overcome to support the scientific objectives of
\CE{}.

\subsection{Model Accuracy}\label{sec:model-accuracy}
Currently, the systematic error in the calibrated strain data in the LIGO
detectors is estimated by using a set of offline measurements of the
opto-mechanical and feedback control responses. These measurements, which are
taken while the detector is not actively observing for gravitational waves (i.e.
``offline''), are well-resolved in frequency but poorly and often unevenly
sampled in time~\cite{sun2021characterization,Sun_2020}. These offline
measurements are later paired with ``online'' measurements, taken while the
detector is actively observing for gravitational waves, that are designed to
track slowly varying time-dependent parameters~\cite{Viets_2018, Tuyenbayev_2016}. Combining measurements of
the closed-loop interferometric optical plant and actuators, time variations of
known parameters, and transfer functions of digital and analog electronics,
results in a frequency- and time-dependent length-response model to differential
length variations from external sources (e.g.~gravitational waves). The
length-response model is compared against measurements of the interferometer
length-response to validate the calibration uncertainty estimate.

The systematic error in the calibration of the optical plant is primarily
limited by analytical model of the opto-mechanical length response~\cite{Sun_2020,sun2021characterization}. The
opto-mechanical length response is determined by the reflectivities of the arm
cavity and signal extraction mirrors, the length of the \FP{} cavities in the
arms, and the length of the signal extraction cavity~\cite{MIZUNO1993273,Abbott_2017}. In aLIGO, the optical
plant model is approximated as a single-pole low-pass filter with a time delay
and a gain (see Figure 3 of Ref.~\cite{Abbott_2017})~\cite{Abbott_2017,Cahillane:2017vkb, Sun_2020,PhysRevD.65.042001,Ward:thesis,Hall:thesis}. However, this simplified approach
will not be sufficient for the \CE{} detectors due to 1.) the breakdown of the
single-pole approximation with the lower free spectral range (FSR) of the longer
\FP{} arms~\cite{LIGO:DCC-T050136}, 2.) effects of SEC on the optical plant response to length~\cite{MIZUNO1993273,Hall_2019,Srivastava_2022,martynov_2019},
3.) the inability to model low frequency effects in the plant response seen in
O3~\cite{Sun_2020}, and 4.) limitations of the long-wavelength approximation~\cite{Rakhmanov_2008,Essick_2017}.

The limitations of the single-pole approximation due to the 37.5\,kHz FSR of the
4\,km \FP{} arms in the LIGO detectors is well-studied~\cite{LIGO:DCC-T050136,LIGO:DCC-T060237,Rakhmanov_2008, Rakhmanov_2002}. The FSR of a \FP{}
cavity is given by~\cite{Rakhmanov_2002}
\begin{equation}\label{eq:fsr}
\textrm{FSR} = \frac{c}{2L_{\textrm{arm}}}\textrm{,}
\end{equation}
where $c$ is the speed of light and $L_{\textrm{arm}}$ is the cavity length. The
single-pole approximation is only valid at low frequencies ($f\ll \textrm{FSR}$)~\cite{Rakhmanov_2008} and
hence is only relatively accurate to within 5\% up to $\sim$6\,kHz in aLIGO. Both
of the \CE{} detector designs under consideration will have a significantly
lower FSR: 3.75\,kHz for the 40\,km version and 7.5\,kHz for the 20\,km~\cite{CE_horizon}. As
such, the approximated optical plant model in use by today's gravitational-wave
detectors will need to be replaced by a fundamentally different approach: one
that more adequately resembles the full interferometer response up to several
kilohertz and accurately accounts for the FSR being closer to and, in some
cases, in the detection band.

Additionally, at frequencies that are low compared to the FSR (the long
wavelength approximation), the gravitational strain antenna response of the two
\FP{} arms experiences very little dependence on the direction to the
gravitational-wave source. At these frequencies, changes in source direction
predominantly translate to gain shifts in the antenna response. However, at
frequencies approaching (and beyond) the FSR, the antenna response's dependence
on source direction plays a much more significant
role~\cite{Essick_2017,Rakhmanov_2008,LIGO:DCC-T050136}. See Rakhmanov, et
al.~\cite{Rakhmanov_2008} and Essick, et al.~\cite{Essick_2017} for more
in-depth discussions of the frequency dependent antenna response of
gravitational wave detectors in cases for which the long wavelength
approximation is not applicable. At 37.5\,kHz, the FSR of the LIGO detectors
high enough in frequency to treat the antenna response in the long wavelength
approximation. This allows LIGO detector operations to estimate gravitational
strain by estimating the ratio of test mass displacement to the mean arm cavity
length in real time without consideration for the direction of the gravitational
wave source~\cite{LIGO:DCC-T970101,Sun_2020}. In contrast, the 20\,km and 40\,km
long arms of the two \CE{} detectors will have FSRs that are lower by an order
of magnitude. More importantly, the \CE{} FSRs and their effects will overlap
with the future detectors' frequency band of interest. At frequencies near the
FSR, gravitational wave source direction must be taken into account in order to
translate test mass displacement into strain. As such, it is likely that the
gravitational strain time series will not be produced in real time at the
detector. Instead, a calibrated differential test mass displacement time series
is the most likely data stream that will be provided to analysis pipelines, each
of which will need to account for source direction as part of their real time
operation.

Another effect that will impact the optical response in the \CE{} detectors is
that of the coupled cavity resonance of the SEC. The placement of a
signal extraction mirror in the interferometer layout results in a coupled
cavity resonance. This resonance is located in frequency at~\cite{martynov_2019,Srivastava_2022}
\begin{equation}\label{eq:sec_resonance}
  f_{s} = \frac{c}{4\pi}\sqrt{\frac{T_{\textrm{i}}}{L_{\textrm{arm}}L_{\textrm{s}}}}
\end{equation}
and has bandwidth
\begin{equation}
  \gamma_{s}= \frac{cT_{\textrm{s}}}{8\pi L_{\textrm{s}}}\textrm{,}
\end{equation}
where $T_{i}$ is the transmissivity of the input test masses, $L_{\textrm{s}}$ is the
length of the SEC, and $T_{s}$ is the transmissivity of the of signal extraction
mirror. The width of the SEC's coupled cavity resonance, $\gamma_{\textrm{s}}$,
is directly proportional to the ratio $T_{\textrm{s}}/L_{\textrm{s}}$. This
effect is significant because a narrow SEC coupled cavity resonance 1.) will
affect the optical response of the interferometer if it lies in or near the
detection band and 2.) can in turn narrow the bandwidth of the interferometer as
a whole.

Without additional corrections, the simple single-pole approximation also fails
to model SEC detuning effects that manifest below 20\,Hz~\cite{Cahillane:2017vkb,Sun_2020}. Poorly modeled
time-dependent optical plant effects have been observed at both LIGO detectors
below 30\,Hz~\cite{Cahillane:2017vkb,Sun_2020,sun2021characterization}. The process that causes these effects, called
``thermalization,'' is further complicated by point-like defects in the optical
coatings of the test masses~\cite{sun2021characterization}. It is reasonable to expect the effects from
thermalization to become more pronounced (and possibly more challenging to
compensate) as gravitational-wave detectors continue to increase the power in
their arm cavities. Moreover, evidence of multiple-input multiple-output (MIMO)
cross-coupling between the DARM loop and the alignment sensing and control loops
surfaced in the first half of O3. This cross-coupling was caused by changes to
the detector alignment scheme that positioned the main interferometer laser beam
away from the center of some of the test mass optics to avoid the most impactful
point defects~\cite{sun2021characterization}.

As the current generation of gravitational-wave detectors continues to evolve,
so too must our techniques to model the interferometers' opto-mechanical
responses. Additional work will be needed to go beyond the approximations in use
today and to characterize the full dual-recycled \FP{} Michelson interferometer
response in such a way that the FSR effects of the \FP{} cavities, SEC detuning,
MIMO cross-coupling effects, and the detector's thermalization process are
accurately modeled and compensated for.

\subsection{Absolute Reference}\label{sec:absol-displ-stand}
The primary reference tool for test mass displacement for all current
gravitational-wave detectors is an independent auxiliary laser system known as
the photon calibrator (Pcal)~\cite{Karki_2016}. The Pcal is used to apply a known periodic
force on the end test mass optics via radiation pressure~\cite{Goetz_2009}. As the fundamental
reference against which all measurements of displacement and gravitational-wave
strain thereof, these fiducial displacements directly determine the overall
absolute calibration of the detector output in response to displacement of the
test masses~\cite{galaxies10020042}. In O3, the fiducial displacements imposed by the Pcal team at
the two LIGO detectors reached sub-percent accuracy with relative uncertainties
of $0.41\%$ at LIGO Hanford and $0.44\%$ at LIGO
Livingston~\cite{sun2021characterization}.

While the Pcal system has seen many significant improvements since its
inception, additional research and development is needed to realize the
$\lesssim 0.1\%$ accuracy expected by \CE{}~\cite{CE_horizon}. This future work will likely need to
focus on the mitigation of various sources of systematic error associated with
the use of the Pcal system, including constraints of the Pcal beam positions on
the test masses relative to the masses' center of inertia as well as to the main
interferometer beam. Offsets in the Pcal beam locations on the end test masses
can introduce undesirable angular motion that can compromise the accuracy of
displacement fiducials~\cite{Goetz:2009wf}. In principle, it's expected that one could model and
subtract the contributions from angular rotation due to the Pcal system. Such a
feat would require precise knowledge of the Pcal and main interferometer beam
locations on the test mass~\cite{galaxies10020042}. Other sources of systematic error that may need
to be addressed include modeling of and accounting for the so-called \emph{local elastic deformation}
effect~\cite{Hild_2007,Goetz_2009}. This effect causes the center of the test mass surface to move along
the axis of the main interferometer laser due to the coupling from the Pcal
force to the resonant modes of the test mass. Moreover, local elastic
deformation effect was demonstrated to produce errors as large as $50\%$ when
operating the Pcal in a single-beam configuration pointed at the center of the
test mass~\cite{Goetz:2009wf}.
Future work to model these effects will need to take into account the
larger and more massive test masses of \CE{}.

Ultimately, the many strengths of using the Pcal system as an absolute reference
depends on the accuracy with which the force being applied by the Pcal laser
onto the detector's test masses is known. Presently, the force being applied is
measured by using a photodiode to monitor the light that is reflected off of the
test mass from the Pcal system's auxiliary laser. The power sensing calibration
of the photodiode and its enclosing apparatus is traceable to SI units via
measurements performed at the National Institute of Standards (NIST) in the
United States and Physikalisch-Technische Bundesanstalt (PTB) in Germany~\cite{Bhattacharjee_2020,galaxies10020042}.
Further improvement in the power sensing accuracy of the Pcal system will
depend, at least in part, on NIST, PTB, and institutes of metrology in other
nations. In fact, recent efforts to establish a global standard-sharing network
have begun to ensure improved power sensing accuracy across multiple
gravitational-wave detectors across the globe. The use of this network, coupled
with ongoing and future efforts to limit systematic error should go a long way
toward realizing less than 0.1\% uncertainty in the absolute displacement
reference of current and future detectors~\cite{galaxies10020042}.

Another calibration standard that has been under development and is currently
being considered is the Newtonian calibrator (Ncal)~\cite{Estevez_2018,Estevez_2021,Ross_2021}. The Ncal system employs
rapidly spinning masses near the detector's optics to induce gravitational
gradient forces on the test masses~\cite{Inoue_2018,Estevez_2018}. Ncal development has primarily taken
place at Virgo Observatory.~\cite{Estevez_2021}. LIGO Hanford Observatory also has a variant of
the Ncal system installed for additional testing~\cite{Ross_2021}. Work to bring the
systematic uncertainty of Ncal system below $1\%$ is
ongoing~\cite{Estevez_2021}. If the Ncal is able to provide continuous,
sub-percent calibration fiducials at the observatories where it is installed, it
could represent an important analog to the calibration provided by the Pcal
system at frequencies where the two overlap. In addition, the Ncal system has
the distinct potential advantage that its use could obviate the need for
in-chamber optical-efficiency measurements that the Pcal relies on to achieve
sub-percent accuracy~\cite{galaxies10020042}.

In addition to calibration via radiation pressure and gravitational gradients,
other calibration standards are being considered for use with future detectors.
These include using existing detector subsystems such as radio-frequency
oscillation of the auxiliary green lasers~\cite{Abbott_2017} and laser-frequency
modulation of the main interferometer laser~\cite{Goetz_2010}. Calibration based
on gravitational-wave sources themselves has also been studied. However, these
methods are not likely to be competitive with Pcal-based calibration in the next
generation of ground-based gravitational-wave
detectors~\cite{Essick_2019,schutz2020selfcalibration}.

\section{Discussion}\label{sec:discuss}

\begin{table*}[p]
    \centering
\begin{tabularx}{\textwidth}{| p{0.24\textwidth} X |}
\hline
    \multicolumn{1}{|c|}{\textbf{Technical goal}}& \multicolumn{1}{c|}{\textbf{Key science objectives realized}} \\
    \hline\hline
    \vspace{2mm}
    Low controls noise & 
    \begin{itemize}
        \item Significant increase in sensitivity to cosmological sources
        \item Significant increase in signal-to-noise ratio of all detected signals
        \item Significant increase in early warning time for CBC signals
    \end{itemize}\\ \hline
    \vspace{2mm}
    Low glitch rate & 
    \begin{itemize}
        \item Increased chance of identification of remnants of Population III stars and PBH
        \item Decreased risk of data quality issues preventing analysis of short, rare events such as SNe
        \item Decreased risk of systematic biases for analyses of astrophysical signals
    \end{itemize}\\ \hline
    \vspace{2mm}
    High uptime & 
    \begin{itemize}
        \item Reduced risk of missing rare events such as SNe, nearby multi-messenger astronomy events, or multi-band events
        \item Decreased uncertainty in sky localization
    \end{itemize}\\ \hline
    \vspace{2mm}
    Low line density & 
    \begin{itemize}
        \item Ability to search large frequency spaces for persistent gravitational waves
    \end{itemize}\\ \hline
    \vspace{1mm}
    Low coherent noise & 
    \begin{itemize}
        \item Increased chance to identify cosmological stochastic background
    \end{itemize}\\ \hline
    \vspace{2mm}
    Low calibration \newline uncertainty & 
    \begin{itemize}
        \item Increased precision in the measurement of the nuclear equation of state at finite and zero temperatures
        \item Improved precision on measurements of the expansion of the universe
        \item Decreased risk of systematic biases for all astrophysical analyses
    \end{itemize}\\ \hline
\end{tabularx}
\caption{Summary of the main science goals that are impacted by the different technical noise sources discussed in this work. A wide range of different science themes will be enabled by improvements in technical noise, data quality, and calibration for XG detectors. Note that this table is not exhaustive and does not contain all of the science objectives that would be realized by addressing these goals.}
\label{tab:XG_sci_summ}
\end{table*}

In this work, we have outlined how the understanding of technical noise, data quality, and calibration will be essential to realize the scientific objectives of XG gravitational-wave detectors. 
A summary of the key science themes that we identified as highly reliant on meeting our goals in these areas is shown in Table~\ref{tab:XG_sci_summ}. As designs for XG detectors are still under development, effects from different types of technical noise in these new detectors have not been extensively studied. When possible, design work should weigh the contribution of technical noises from various design parameters, especially if those noises have the potential to significantly inhibit science goals.

It is important to remind the reader that the majority of the projections and discussions in this work are based on the current state of technical noise, data quality, and calibration, and are not the full picture of the impact these challenges will have on XG science. 
Considering the improvements made in the last decade in gravitational-wave science, it is perhaps expected that significant further research will yield improvements in all of these areas by the XG detector era.
For example, comparing the Enhanced LIGO noise budget~\cite{LIGOScientific:2007fwp} to the Advanced LIGO budget~\cite{o3LIGOinstrumentation}, there was over an order of magnitude improvement in the level of controls noise for both length and alignment controls.
Similarly, we have seen orders of magnitude improvements in the glitch rate between observing runs, as was the case for Advanced Virgo between O2 and O3~\cite{GWTC-3}.
And finally, for calibration, the LIGO Livingston calibration uncertainty in 2010 was up to 18\% at some frequencies~\cite{Bartos:2011aaa}, while in O2, the uncertainty over the same frequency band had dropped to 3\%~\cite{Cahillane:2017vkb}.
Therefore, the conclusions in this work should be interpreted as a reminder that continued progress in technical noise, data quality, and calibration is integral to achieving XG science goals rather than suggesting that these challenges are insurmountable. Even while XG detectors remain in the design phase, current generation gravitational wave detectors continue to expand in their abilities; additional research and development on techniques to improve the sensitivity of detectors and their data quality, and better defining methods to improve on calibration will be essential to maximize the science potential of both the current and next generation of gravitational-wave astrophysics.

\section{Acknowledgments}

The authors thank Ansel Neunzert and Evan Goetz for helpful discussions
regarding narrow-band noise features. The authors also thank Stefan Ballmer,
Craig Cahillane, Kevin Kuns, Evan Hall, and Matt Evans for helpful discussions and suggestions
regarding technical noise and interferometer design choices. We also thank
Joshua Smith, Jocelyn Read, and Alex Nitz for their comments and suggestions regarding \CE{} science
projections and Joseph Betzwieser, Jeff Kissel, Kevin Kuns, Richard Savage, Ling
Sun, Vlad Bossilkov, Keita Kawabe, Jameson Rollins, and Evan Goetz for informative
discussions and comments about calibration in the XG era.
EC is supported by NSF awards PHY-2207640 and AST-2219109.
LD is supported by NSF award PHY-1912598.
DD is supported by the NSF through the LIGO Laboratory.

This material is based upon work supported by NSF’s LIGO Laboratory 
which is a major facility fully funded by the 
National Science Foundation.
LIGO was constructed by the California Institute of Technology 
and Massachusetts Institute of Technology with funding from 
the National Science Foundation, 
and operates under cooperative agreement PHY-1764464. 
Advanced LIGO was built under award PHY-0823459.
The authors are grateful for computational resources provided by the 
LIGO Laboratory and supported by 
National Science Foundation Grants PHY-0757058 and PHY-0823459.
This work carries Cosmic Explorer document number CE-P2400001.

\appendix
\section{Derivation of glitch impact values}
\label{app:glitch_rate}

To evaluate the effect of varying glitch rates, we take advantage of information about search sensitivity, detector sensitivity, and glitch rates from GWTC-3~\cite{GWTC-3}.
Our goal is to model the sensitivity of CBC search pipelines as a combination of both the detector sensitivity and the rate of glitches in the detector. 
For this purpose, we consider the PyCBC~\cite{Usman:2015kfa} and cWB~\cite{Klimenko:2008fu} search pipelines.
These two are chosen as they split the entire observing period into multiple chunks and analyze each time period separately.
This allows us to better measure the varying sensitivity resulting from glitches compared to pipelines with only a single background model that includes data from the entire observing period. 

We first must estimate the rate of glitches in each short chunk of data. 
We use the rate of glitches for LIGO Hanford and LIGO Livingston during O3 that was released alongside GWTC-3~\cite{GWTC3_figures}. 
We then assign each chunk, $N$, a single value for the rate of glitches, $R_{\text{G},N}$, by evaluating the median value of the sum of the two individual detector glitch rates. 
Therefore, $R_{\text{G},N}$ can be written as
\begin{equation}
    R_{\text{G},N} = \text{med}\{ R_{\text{G},H}(t) + R_{\text{G},L}(t) \} ,
\end{equation}
where $R_{\text{G},H}(t)$ and $R_{\text{G},L}(t)$ are the time-dependent glitch rates for the LIGO Hanford and LIGO Livingston detectors during the chunk in question. 
For this analysis, we do not consider the glitch rate in the Virgo detector, as we assume that the sensitivity of the LIGO-Virgo-KAGRA network in O3 was driven by the sensitivities of the two LIGO detectors. 

We next model the expected sensitivity of the detector network. 
This is directly taken from Figure~1 of GWTC-3~\cite{GWTC-3,GWTC3_figures} that contains the sensitive volume-time (VT) of the LIGO-Virgo-KAGRA detector network during the advanced detector era. 
Specifically, this VT estimate relies only on the BNS inspiral range~\cite{Chen:2017wpg} of each detector in the network.
We denote this VT estimate for each chunk, $N$, as $\text{VT}_{\text{R},N}$.

Finally, we estimate the actual sensitivity of the search pipelines from analyses of simulated compact binary signals released along GWTC-3~\cite{GWTC3_search_sens}.
To focus on the short-duration, redshifted signals that would be detected at the horizon distance of \CE, we only consider injections with a total detector-frame mass greater than $50M_\odot$.
For this rough estimate, we estimate the total VT per chunk as the total number of injections recovered with an inverse false alarm rate greater than one year. 
We label the sensitivity based on these injections per chunk as $\text{VT}_{\text{I},N}$.

With these values computed, we can now approximate the injection-based VT ($\text{VT}_{\text{I},N}$) as a combination of the range-based VT ($\text{VT}_{\text{R},N}$) and the glitch rate ($R_{\text{G},N}$).
More precisely, we assume differences between the range-based VT and the injection-based VT are due to the relative glitch rate for each chunk. 
We also allow the glitch rate  to contribute to the total VT with an exponent, $\alpha$. 
Hence, we relate these quantities via
\begin{equation} \label{eq:glitch_vt}
    \left( \frac{\text{VT}_{\text{I},N}}{\text{VT}_{\text{R},N}} \right) 
    \text{med}\left\{ \frac{\text{VT}_{\text{R},i}}{\text{VT}_{\text{I},i}} \right\} 
    = \left( \frac{R_{\text{G},N}}{\text{med}\{R_{\text{G},i}\}} \right)^\alpha .
\end{equation}
The value of $\alpha$ must be fit based on our available data. 
For PyCBC, we estimate $\alpha=0.05$ and for cWB, we estimate $\alpha=0.10$.
From this result, we can estimate how the sensitivity of the search pipelines would change with respect to the glitch rate in the detectors. 

While this model is insufficient to explain all of the differences between the range-based VT and the injection-based VT, we do find that the glitch rate can account for some of the variation. 
Using the Pearson correlation coefficient~\cite{Pearson_corr}, we estimate that the two sides of Equation~\ref{eq:glitch_vt} are linearly correlated with a p-value of $9 \times 10^{-3}$ in the case of PyCBC and $4 \times 10^{-4}$ in the case of cWB. 

As we have done this calculation considering the glitch rate relative to the median for O3, we also need to consider the absolute loss in sensitivity from an O3-like glitch rate. 
To do this, we use the same injections used to estimate $\text{VT}_{\text{I},N}$. 
We find that at $\text{SNR}\approx 10$, 50\% of all BNS injections are recovered by PyCBC. 
We define this SNR threshold as the ``ideal sensitivity.'' 
For the masses considered in this calculation, the SNR threshold for 50\% detection efficiency is $\text{SNR}\approx 11$ for PyCBC and $\text{SNR}\approx 13$ for cWB. 
Hence there is a 9\% loss in sensitivity for PyCBC and 23\% loss in sensitivity for PyCBC. 

A more complete estimate of how glitch rates could impact detection efficiency could be done by comparing real data to simulated Gaussian noise. 
Such an analysis would be difficult to fully account for all of the confounding variables, such as the variable sensitivities of the detector and the varying duty cycles across the network. 
However, given the significant loss in sensitivity that can be attributed to glitches in this analysis, further analysis is warranted.

\section*{References}
\bibliographystyle{my-iopart-num}
\bibliography{main.bbl}

\providecommand{\newblock}{}
\providecommand{\url}[1]{{\tt #1}}
\providecommand{\urlprefix}{}
\providecommand{\href}[2]{#2}
\begin{thebibliography}{100}

\bibitem{GW150914_paper}
{Abbott} B~P, {Abbott} R, {Abbott} T~D, {Abernathy} M~R {\em et~al.\/} 2016
  \href{https://doi.org/10.1103/PhysRevLett.116.061102}{{\em Physical Review
  Letters\/} {\bf 116} 061102} (\textit{Preprint}
  \href{https://arxiv.org/abs/1602.03837}{{\tt 1602.03837}})

\bibitem{GW170817}
Abbott B~P {\em et~al.\/} (LIGO Scientific, Virgo) 2017
  \href{https://doi.org/10.1103/PhysRevLett.119.161101}{{\em Phys. Rev.
  Lett.\/} {\bf 119} 161101} (\textit{Preprint}
  \href{https://arxiv.org/abs/1710.05832}{{\tt 1710.05832}})

\bibitem{GWTC-1}
Abbott B~P {\em et~al.\/} (LIGO Scientific, Virgo) 2019
  \href{https://doi.org/10.1103/PhysRevX.9.031040}{{\em Phys. Rev. X\/}
  031040} (\textit{Preprint} \href{https://arxiv.org/abs/1811.12907}{{\tt
  1811.12907}})

\bibitem{GWTC-2}
Abbott R {\em et~al.\/} (LIGO Scientific, Virgo) 2021
  \href{https://doi.org/10.1103/PhysRevX.11.021053}{{\em Phys. Rev. X\/} {\bf
  11} 021053} (\textit{Preprint} \href{https://arxiv.org/abs/2010.14527}{{\tt
  2010.14527}})

\bibitem{GWTC-2.1}
Abbott R {\em et~al.\/} (LIGO Scientific, Virgo) 2021 ArXiv:2108.01045
  \textit{Preprint} \href{https://arxiv.org/abs/2108.01045}{{\tt 2108.01045}}

\bibitem{GWTC-3}
Abbott R {\em et~al.\/} (KAGRA, Virgo, LIGO Scientific) 2023
  \href{https://doi.org/10.1103/PhysRevX.13.041039}{{\em Phys. Rev. X\/} {\bf
  13} 041039} (\textit{Preprint} \href{https://arxiv.org/abs/2111.03606}{{\tt
  2111.03606}})

\bibitem{aligo}
Aasi J {\em et~al.\/} (LIGO Scientific) 2015
  \href{https://doi.org/10.1088/0264-9381/32/7/074001}{{\em Class. Quant.
  Grav.\/} {\bf 32} 074001} (\textit{Preprint}
  \href{https://arxiv.org/abs/1411.4547}{{\tt 1411.4547}})

\bibitem{avirgo}
Acernese F {\em et~al.\/} (Virgo) 2015
  \href{https://doi.org/10.1088/0264-9381/32/2/024001}{{\em Class. Quant.
  Grav.\/} {\bf 32} 024001} (\textit{Preprint}
  \href{https://arxiv.org/abs/1408.3978}{{\tt 1408.3978}})

\bibitem{kagra}
Akutsu T {\em et~al.\/} (KAGRA) 2019
  \href{https://doi.org/10.1038/s41550-018-0658-y}{{\em Nature Astron.\/} {\bf
  3} 35--40} (\textit{Preprint} \href{https://arxiv.org/abs/1811.08079}{{\tt
  1811.08079}})

\bibitem{o3LIGOinstrumentation}
Buikema A {\em et~al.\/} (LIGO Scientific) 2020
  \href{https://doi.org/10.1103/PhysRevD.102.062003}{{\em Phys. Rev. D\/} {\bf
  102} 062003} (\textit{Preprint} \href{https://arxiv.org/abs/2008.01301}{{\tt
  2008.01301}})

\bibitem{Virgo_O3_detchar}
Acernese F {\em et~al.\/} (Virgo) 2023
  \href{https://doi.org/10.1088/1361-6382/acd92d}{{\em Class. Quant. Grav.\/}
  {\bf 40} 185006} (\textit{Preprint}
  \href{https://arxiv.org/abs/2210.15633}{{\tt 2210.15633}})

\bibitem{KAGRA_O3GK_performance}
Abe H {\em et~al.\/} (KAGRA) 2023
  \href{https://doi.org/10.1093/ptep/ptac093}{{\em PTEP\/} {\bf 2023} 10A101}
  (\textit{Preprint} \href{https://arxiv.org/abs/2203.07011}{{\tt 2203.07011}})

\bibitem{KAGRA_detchar}
Akutsu T {\em et~al.\/} (KAGRA) 2021
  \href{https://doi.org/10.1093/ptep/ptab018}{{\em PTEP\/} {\bf 2021} 05A102}
  (\textit{Preprint} \href{https://arxiv.org/abs/2009.09305}{{\tt 2009.09305}})

\bibitem{CE_horizon}
Evans M {\em et~al.\/} 2021 \textit{Preprint}
  \href{https://arxiv.org/abs/2109.09882}{{\tt 2109.09882}}

\bibitem{ET_design}
Punturo M {\em et~al.\/} 2010
  \href{https://doi.org/10.1088/0264-9381/27/19/194002}{{\em Class. Quant.
  Grav.\/} {\bf 27} 194002}

\bibitem{NEMO}
Ackley K {\em et~al.\/} 2020 \href{https://doi.org/10.1017/pasa.2020.39}{{\em
  Publ. Astron. Soc. Austral.\/} {\bf 37} e047} (\textit{Preprint}
  \href{https://arxiv.org/abs/2007.03128}{{\tt 2007.03128}})

\bibitem{CE_LF}
Hall E~D {\em et~al.\/} 2021
  \href{https://doi.org/10.1103/PhysRevD.103.122004}{{\em Phys. Rev. D\/} {\bf
  103} 122004} (\textit{Preprint} \href{https://arxiv.org/abs/2012.03608}{{\tt
  2012.03608}})

\bibitem{GW150914_detchar}
Abbott B~P {\em et~al.\/} (LIGO Scientific, Virgo) 2016
  \href{https://doi.org/10.1088/0264-9381/33/13/134001}{{\em Class. Quant.
  Grav.\/} {\bf 33} 134001} (\textit{Preprint}
  \href{https://arxiv.org/abs/1602.03844}{{\tt 1602.03844}})

\bibitem{Nuttall:2018xhi}
Nuttall L~K 2018 \href{https://doi.org/10.1098/rsta.2017.0286}{{\em Phil.
  Trans. Roy. Soc. Lond.\/} {\bf A376} 20170286} (\textit{Preprint}
  \href{https://arxiv.org/abs/1804.07592}{{\tt 1804.07592}})

\bibitem{LIGO_O3_detchar}
Davis D {\em et~al.\/} (LIGO Scientific) 2021
  \href{https://doi.org/10.1088/1361-6382/abfd85}{{\em Class. Quant. Grav.\/}
  {\bf 38} 135014} (\textit{Preprint}
  \href{https://arxiv.org/abs/2101.11673}{{\tt 2101.11673}})

\bibitem{LIGOScientific:2019hgc}
Abbott B~P {\em et~al.\/} (LIGO Scientific, Virgo) 2020
  \href{https://doi.org/10.1088/1361-6382/ab685e}{{\em Class. Quant. Grav.\/}
  {\bf 37} 055002} (\textit{Preprint}
  \href{https://arxiv.org/abs/1908.11170}{{\tt 1908.11170}})

\bibitem{Davis:2022dnd}
Davis D and Walker M 2022 \href{https://doi.org/10.3390/galaxies10010012}{{\em
  Galaxies\/} {\bf 10} 12}

\bibitem{TheLIGOScientific:2017lwt}
Abbott B~P {\em et~al.\/} (LIGO Scientific, Virgo) 2018
  \href{https://doi.org/10.1088/1361-6382/aaaafa}{{\em Class. Quant. Grav.\/}
  {\bf 35} 065010} (\textit{Preprint}
  \href{https://arxiv.org/abs/1710.02185}{{\tt 1710.02185}})

\bibitem{Powell:2018csz}
Powell J 2018 \href{https://doi.org/10.1088/1361-6382/aacf18}{{\em Class.
  Quant. Grav.\/} {\bf 35} 155017} (\textit{Preprint}
  \href{https://arxiv.org/abs/1803.11346}{{\tt 1803.11346}})

\bibitem{Kwok:2021zny}
Kwok J~Y~L, Lo R~K~L, Weinstein A~J and Li T~G~F 2022
  \href{https://doi.org/10.1103/PhysRevD.105.024066}{{\em Phys. Rev. D\/} {\bf
  105} 024066} (\textit{Preprint} \href{https://arxiv.org/abs/2109.07642}{{\tt
  2109.07642}})

\bibitem{Mozzon:2021wam}
Mozzon S, Ashton G, Nuttall L~K and Williamson A~R 2022
  \href{https://doi.org/10.1103/PhysRevD.106.043504}{{\em Phys. Rev. D\/} {\bf
  106} 043504} (\textit{Preprint} \href{https://arxiv.org/abs/2110.11731}{{\tt
  2110.11731}})

\bibitem{Macas:2022afm}
Macas R, Pooley J, Nuttall L~K, Davis D {\em et~al.\/} 2022
  \href{https://doi.org/10.1103/PhysRevD.105.103021}{{\em Phys. Rev. D\/} {\bf
  105} 103021} (\textit{Preprint} \href{https://arxiv.org/abs/2202.00344}{{\tt
  2202.00344}})

\bibitem{Hourihane:2022doe}
Hourihane S, Chatziioannou K, Wijngaarden M, Davis D {\em et~al.\/} 2022
  \href{https://doi.org/10.1103/PhysRevD.106.042006}{{\em Phys. Rev. D\/} {\bf
  106} 042006} (\textit{Preprint} \href{https://arxiv.org/abs/2205.13580}{{\tt
  2205.13580}})

\bibitem{Ghonge:2023ksb}
Ghonge S, Brandt J, Sullivan J~M, Millhouse M {\em et~al.\/} 2023
  \textit{Preprint} \href{https://arxiv.org/abs/2311.09159}{{\tt 2311.09159}}

\bibitem{Davis:2022ird}
Davis D, Littenberg T~B, Romero-Shaw I~M, Millhouse M {\em et~al.\/} 2022
  \href{https://doi.org/10.1088/1361-6382/aca238}{{\em Class. Quant. Grav.\/}
  {\bf 39} 245013} (\textit{Preprint}
  \href{https://arxiv.org/abs/2207.03429}{{\tt 2207.03429}})

\bibitem{2023ApPhL.122r4101B}
{Berger} B~K {\em et~al.\/} 2023 \href{https://doi.org/10.1063/5.0140766}{{\em
  Applied Physics Letters\/} {\bf 122} 184101}

\bibitem{Sun_2020}
Sun L {\em et~al.\/} 2020 \href{https://doi.org/10.1088/1361-6382/abb14e}{{\em
  Class. Quant. Grav.\/} {\bf 37} 225008} (\textit{Preprint}
  \href{https://arxiv.org/abs/2005.02531}{{\tt 2005.02531}})

\bibitem{Abbott_2017}
Abbott B~P {\em et~al.\/} (LIGO Scientific) 2017
  \href{https://doi.org/10.1103/PhysRevD.95.062003}{{\em Phys. Rev. D\/} {\bf
  95} 062003} (\textit{Preprint} \href{https://arxiv.org/abs/1602.03845}{{\tt
  1602.03845}})

\bibitem{Essick_2022}
Essick R 2022 \href{https://doi.org/10.1103/PhysRevD.105.082002}{{\em Phys.
  Rev. D\/} {\bf 105} 082002} (\textit{Preprint}
  \href{https://arxiv.org/abs/2202.00823}{{\tt 2202.00823}})

\bibitem{Cahillane:2017vkb}
Cahillane C {\em et~al.\/} (LIGO Scientific) 2017
  \href{https://doi.org/10.1103/PhysRevD.96.102001}{{\em Phys. Rev.\/} {\bf
  D96} 102001} (\textit{Preprint} \href{https://arxiv.org/abs/1708.03023}{{\tt
  1708.03023}})

\bibitem{Virgo:Cal_O2_Acernese_2018}
Acernese F, Adams T, Agatsuma K, Aiello L {\em et~al.\/} 2018
  \href{https://doi.org/10.1088/1361-6382/aadf1a}{{\em Classical and Quantum
  Gravity\/} {\bf 35} 205004}
  \urlprefix\url{http://dx.doi.org/10.1088/1361-6382/aadf1a}

\bibitem{Karki_2016}
{Karki} S, {Tuyenbayev} D, {Kandhasamy} S, {Abbott} B~P {\em et~al.\/} 2016
  \href{https://doi.org/10.1063/1.4967303}{{\em Review of Scientific
  Instruments\/} {\bf 87} 114503} (\textit{Preprint}
  \href{https://arxiv.org/abs/1608.05055}{{\tt 1608.05055}})

\bibitem{Bhattacharjee_2020}
Bhattacharjee D, Lecoeuche Y, Karki S, Betzwieser J {\em et~al.\/} 2021
  \href{https://doi.org/10.1088/1361-6382/aba9ed}{{\em Class. Quant. Grav.\/}
  {\bf 38} 015009} (\textit{Preprint}
  \href{https://arxiv.org/abs/2006.00130}{{\tt 2006.00130}})

\bibitem{galaxies10020042}
Karki S, Bhattacharjee D and Savage R~L 2022
  \href{https://doi.org/10.3390/galaxies10020042}{{\em Galaxies\/} {\bf 10} 42}

\bibitem{o1detectorpaper}
Abbott B~P {\em et~al.\/} 2016
  \href{https://doi.org/10.1103/PhysRevD.93.112004}{{\em Phys. Rev. D\/} {\bf
  93} 112004} [Addendum: Phys.Rev.D 97, 059901 (2018)] (\textit{Preprint}
  \href{https://arxiv.org/abs/1604.00439}{{\tt 1604.00439}})

\bibitem{KAGRA:2022fgc}
Abe H {\em et~al.\/} (KAGRA) 2023
  \href{https://doi.org/10.1093/ptep/ptac093}{{\em PTEP\/} {\bf 2023} 10A101}
  (\textit{Preprint} \href{https://arxiv.org/abs/2203.07011}{{\tt 2203.07011}})

\bibitem{Srivastava_2022}
Srivastava V, Davis D, Kuns K, Landry P {\em et~al.\/} 2022
  \href{https://doi.org/10.3847/1538-4357/ac5f04}{{\em Astrophys. J.\/} {\bf
  931} 22} (\textit{Preprint} \href{https://arxiv.org/abs/2201.10668}{{\tt
  2201.10668}})

\bibitem{Meadors:2013lja}
Meadors G~D, Kawabe K and Riles K 2014
  \href{https://doi.org/10.1088/0264-9381/31/10/105014}{{\em Class. Quant.
  Grav.\/} {\bf 31} 105014} (\textit{Preprint}
  \href{https://arxiv.org/abs/1311.6835}{{\tt 1311.6835}})

\bibitem{CooperHoQI}
Cooper S~J, Green A~C, Collins C~J, Hoyland D {\em et~al.\/} 2018
  \href{https://doi.org/10.1088/1361-6382/aab2e9}{{\em Class. Quant. Grav.\/}
  {\bf 35} 095007} (\textit{Preprint}
  \href{https://arxiv.org/abs/1710.05943}{{\tt 1710.05943}})

\bibitem{BarsottiASC}
Barsotti L, Evans M and Fritschel P 2010
  \href{https://doi.org/10.1088/0264-9381/27/8/084026}{{\em Class. Quant.
  Grav.\/} {\bf 27} 084026}

\bibitem{SidlesSigg}
Sidles J~A and Sigg D 2006
  \href{https://doi.org/10.1016/j.physleta.2006.01.051}{{\em Phys. Lett. A\/}
  {\bf 354} 167--72}

\bibitem{ASC_MarieK}
Kasprzack M 2021 {ASC} noise budgeting for {LLO} Tech. Rep. DCC-G2100751 {LIGO}
  \urlprefix\url{https://dcc.ligo.org/LIGO-G2100751/public}

\bibitem{Maggiore:2024eun}
Maggiore R {\em et~al.\/} 2024 \textit{Preprint}
  \href{https://arxiv.org/abs/2401.13013}{{\tt 2401.13013}}

\bibitem{CE_noise_curve}
Kuns K {\em et~al.\/} 2020 {Cosmic Explorer} strain sensitivity Tech. Rep.
  CE-T2000017 {Cosmic Explorer}
  \urlprefix\url{https://dcc.cosmicexplorer.org/CE-T2000017/public}

\bibitem{Chen:2017wpg}
Chen H~Y, Holz D~E, Miller J, Evans M {\em et~al.\/} 2021
  \href{https://doi.org/10.1088/1361-6382/abd594}{{\em Class. Quant. Grav.\/}
  {\bf 38} 055010} (\textit{Preprint}
  \href{https://arxiv.org/abs/1709.08079}{{\tt 1709.08079}})

\bibitem{inspiral_range}
Rollins J and Creighton J {inspiral-range: version 0.9.1}
  \url{https://git.ligo.org/gwinc/inspiral-range/tree/0.9.1}

\bibitem{Pratten:2020ceb}
Pratten G {\em et~al.\/} 2021
  \href{https://doi.org/10.1103/PhysRevD.103.104056}{{\em Phys. Rev. D\/} {\bf
  103} 104056} (\textit{Preprint} \href{https://arxiv.org/abs/2004.06503}{{\tt
  2004.06503}})

\bibitem{Planck:2018vyg}
Aghanim N {\em et~al.\/} (Planck) 2020
  \href{https://doi.org/10.1051/0004-6361/201833910}{{\em Astron. Astrophys.\/}
  {\bf 641} A6} [Erratum: Astron.Astrophys. 652, C4 (2021)] (\textit{Preprint}
  \href{https://arxiv.org/abs/1807.06209}{{\tt 1807.06209}})

\bibitem{Damour:2012yf}
Damour T, Nagar A and Villain L 2012
  \href{https://doi.org/10.1103/PhysRevD.85.123007}{{\em Phys. Rev. D\/} {\bf
  85} 123007} (\textit{Preprint} \href{https://arxiv.org/abs/1203.4352}{{\tt
  1203.4352}})

\bibitem{Chan:2018csa}
Chan M~L, Messenger C, Heng I~S and Hendry M 2018
  \href{https://doi.org/10.1103/PhysRevD.97.123014}{{\em Phys. Rev. D\/} {\bf
  97} 123014} (\textit{Preprint} \href{https://arxiv.org/abs/1803.09680}{{\tt
  1803.09680}})

\bibitem{Li:2021mbo}
Li Y, Heng I~S, Chan M~L, Messenger C and Fan X 2022
  \href{https://doi.org/10.1103/PhysRevD.105.043010}{{\em Phys. Rev. D\/} {\bf
  105} 043010} (\textit{Preprint} \href{https://arxiv.org/abs/2109.07389}{{\tt
  2109.07389}})

\bibitem{Borhanian:2022czq}
Borhanian S and Sathyaprakash B~S 2022 \textit{Preprint}
  \href{https://arxiv.org/abs/2202.11048}{{\tt 2202.11048}}

\bibitem{Sathyaprakash:1991mt}
Sathyaprakash B~S and Dhurandhar S~V 1991
  \href{https://doi.org/10.1103/PhysRevD.44.3819}{{\em Phys. Rev. D\/} {\bf 44}
  3819--34}

\bibitem{Grant:2022bla}
Grant A~M and Nichols D~A 2023
  \href{https://doi.org/10.1103/PhysRevD.107.064056}{{\em Phys. Rev. D\/} {\bf
  107} 064056} [Erratum: Phys.Rev.D 108, 029901 (2023)] (\textit{Preprint}
  \href{https://arxiv.org/abs/2210.16266}{{\tt 2210.16266}})

\bibitem{Lenon:2021zac}
Lenon A~K, Brown D~A and Nitz A~H 2021
  \href{https://doi.org/10.1103/PhysRevD.104.063011}{{\em Phys. Rev. D\/} {\bf
  104} 063011} (\textit{Preprint} \href{https://arxiv.org/abs/2103.14088}{{\tt
  2103.14088}})

\bibitem{Saini:2023wdk}
Saini P 2023 \textit{Preprint} \href{https://arxiv.org/abs/2308.07565}{{\tt
  2308.07565}}

\bibitem{KAGRA:2022twx}
Abbott R {\em et~al.\/} (KAGRA, Virgo, LIGO Scientific) 2022
  \href{https://doi.org/10.1093/ptep/ptac073}{{\em PTEP\/} {\bf 2022} 063F01}
  (\textit{Preprint} \href{https://arxiv.org/abs/2203.01270}{{\tt 2203.01270}})

\bibitem{Accadia:2010zzb}
Accadia T {\em et~al.\/} 2010
  \href{https://doi.org/10.1088/0264-9381/27/19/194011}{{\em Class. Quant.
  Grav.\/} {\bf 27} 194011}

\bibitem{Soni:2020rbu}
Soni S {\em et~al.\/} (LIGO Scientific) 2020
  \href{https://doi.org/10.1088/1361-6382/abc906}{{\em Class. Quant. Grav.\/}
  {\bf 38} 025016} (\textit{Preprint}
  \href{https://arxiv.org/abs/2007.14876}{{\tt 2007.14876}})

\bibitem{AdvLIGO:2021oxw}
Nguyen P {\em et~al.\/} (AdvLIGO) 2021
  \href{https://doi.org/10.1088/1361-6382/ac011a}{{\em Class. Quant. Grav.\/}
  {\bf 38} 145001} (\textit{Preprint}
  \href{https://arxiv.org/abs/2101.09935}{{\tt 2101.09935}})

\bibitem{Virgo:2022ypn}
Acernese F {\em et~al.\/} (Virgo) 2022
  \href{https://doi.org/10.1088/1361-6382/ac776a}{{\em Class. Quant. Grav.\/}
  {\bf 39} 235009} (\textit{Preprint}
  \href{https://arxiv.org/abs/2203.04014}{{\tt 2203.04014}})

\bibitem{Soni:2023kqq}
Soni S, Glanzer J, Effler A, Frolov V {\em et~al.\/} 2023 \textit{Preprint}
  \href{https://arxiv.org/abs/2311.05730}{{\tt 2311.05730}}

\bibitem{Washimi:2021ogz}
Washimi T, Yokozawa T, Nakano M, Tanaka T {\em et~al.\/} 2021
  \href{https://doi.org/10.1088/1748-0221/16/07/P07033}{{\em JINST\/} {\bf 16}
  P07033} (\textit{Preprint} \href{https://arxiv.org/abs/2103.06516}{{\tt
  2103.06516}})

\bibitem{Janssens:2022tdj}
Janssens K {\em et~al.\/} 2023
  \href{https://doi.org/10.1103/PhysRevD.107.022004}{{\em Phys. Rev. D\/} {\bf
  107} 022004} (\textit{Preprint} \href{https://arxiv.org/abs/2209.00284}{{\tt
  2209.00284}})

\bibitem{Allen:2005fk}
Allen B, Anderson W~G, Brady P~R, Brown D~A and Creighton J~D~E 2012
  \href{https://doi.org/10.1103/PhysRevD.85.122006}{{\em Phys. Rev. D\/} {\bf
  85} 122006} (\textit{Preprint} \href{https://arxiv.org/abs/0509116}{{\tt
  0509116}})

\bibitem{Nitz:2017lco}
Nitz A~H 2018 \href{https://doi.org/10.1088/1361-6382/aaa13d}{{\em Class.
  Quant. Grav.\/} {\bf 35} 035016} (\textit{Preprint}
  \href{https://arxiv.org/abs/1709.08974}{{\tt 1709.08974}})

\bibitem{Mozzon:2020gwa}
Mozzon S, Nuttall L~K, Lundgren A, Dent T {\em et~al.\/} 2020
  \href{https://doi.org/10.1088/1361-6382/abac6c}{{\em Class. Quant. Grav.\/}
  {\bf 37} 215014} (\textit{Preprint}
  \href{https://arxiv.org/abs/2002.09407}{{\tt 2002.09407}})

\bibitem{Messick:2016aqy}
Messick C {\em et~al.\/} 2017
  \href{https://doi.org/10.1103/PhysRevD.95.042001}{{\em \prd\/} {\bf 95}
  042001} (\textit{Preprint} \href{https://arxiv.org/abs/1604.04324}{{\tt
  1604.04324}})

\bibitem{Zackay:2019kkv}
Zackay B, Venumadhav T, Roulet J, Dai L and Zaldarriaga M 2021
  \href{https://doi.org/10.1103/PhysRevD.104.063034}{{\em Phys. Rev. D\/} {\bf
  104} 063034} (\textit{Preprint} \href{https://arxiv.org/abs/1908.05644}{{\tt
  1908.05644}})

\bibitem{Usman:2015kfa}
Usman S~A {\em et~al.\/} 2016
  \href{https://doi.org/10.1088/0264-9381/33/21/215004}{{\em Class. Quant.
  Grav.\/} {\bf 33} 215004} (\textit{Preprint}
  \href{https://arxiv.org/abs/1508.02357}{{\tt 1508.02357}})

\bibitem{Klimenko:2008fu}
Klimenko S, Yakushin I, Mercer A and Mitselmakher G 2008
  \href{https://doi.org/10.1088/0264-9381/25/11/114029}{{\em Class. Quant.
  Grav.\/} {\bf 25} 114029} (\textit{Preprint}
  \href{https://arxiv.org/abs/0802.3232}{{\tt 0802.3232}})

\bibitem{O2-DQ-paper}
Collaboration L~S 2018 In Preparation

\bibitem{Godwin:2020weu}
Godwin P {\em et~al.\/} 2020 \textit{Preprint}
  \href{https://arxiv.org/abs/2010.15282}{{\tt 2010.15282}}

\bibitem{Beauville:2005au}
Beauville F {\em et~al.\/} 2006
  \href{https://doi.org/10.1088/1742-6596/32/1/032}{{\em J. Phys. Conf. Ser.\/}
  {\bf 32} 212} (\textit{Preprint}
  \href{https://arxiv.org/abs/gr-qc/0509041}{{\tt gr-qc/0509041}})

\bibitem{Goncharov:2022dgl}
Goncharov B, Nitz A~H and Harms J 2022
  \href{https://doi.org/10.1103/PhysRevD.105.122007}{{\em Phys. Rev. D\/} {\bf
  105} 122007} (\textit{Preprint} \href{https://arxiv.org/abs/2204.08533}{{\tt
  2204.08533}})

\bibitem{O2_IMBH}
Abbott B~P {\em et~al.\/} (LIGO Scientific, Virgo) 2019
  \href{https://doi.org/10.1103/PhysRevD.100.064064}{{\em Phys. Rev. D\/} {\bf
  100} 064064} (\textit{Preprint} \href{https://arxiv.org/abs/1906.08000}{{\tt
  1906.08000}})

\bibitem{Payne:2022}
Payne E, Hourihane S, Golomb J, Udall R {\em et~al.\/} 2022
  \href{https://doi.org/10.1103/PhysRevD.106.104017}{{\em Phys. Rev. D\/} {\bf
  106} 104017} (\textit{Preprint} \href{https://arxiv.org/abs/2206.11932}{{\tt
  2206.11932}})

\bibitem{Macas:2023wiw}
Macas R, Lundgren A and Ashton G 2023 \textit{Preprint}
  \href{https://arxiv.org/abs/2311.09921}{{\tt 2311.09921}}

\bibitem{Torres-Forne:2020eax}
Torres-Forn\'e A, Cuoco E, Font J~A and Marquina A 2020
  \href{https://doi.org/10.1103/PhysRevD.102.023011}{{\em Phys. Rev. D\/} {\bf
  102} 023011} (\textit{Preprint} \href{https://arxiv.org/abs/2002.11668}{{\tt
  2002.11668}})

\bibitem{Merritt:2021xwh}
Merritt J, Farr B, Hur R, Edelman and Doctor Z 2021
  \href{https://doi.org/10.1103/PhysRevD.104.102004}{{\em Phys. Rev. D\/} {\bf
  104} 102004} (\textit{Preprint} \href{https://arxiv.org/abs/2108.12044}{{\tt
  2108.12044}})

\bibitem{Udall:2022vkv}
Udall R and Davis D 2023 \href{https://doi.org/10.1063/5.0136896}{{\em Appl.
  Phys. Lett.\/} {\bf 122} 094103} (\textit{Preprint}
  \href{https://arxiv.org/abs/2211.15867}{{\tt 2211.15867}})

\bibitem{Tolley:2023umc}
Tolley A~E, Cabourn~Davies G~S, Harry I~W and Lundgren A~P 2023
  \href{https://doi.org/10.1088/1361-6382/ace22f}{{\em Class. Quant. Grav.\/}
  {\bf 40} 165005} (\textit{Preprint}
  \href{https://arxiv.org/abs/2301.10491}{{\tt 2301.10491}})

\bibitem{Bondarescu:2023jcx}
Bondarescu R, Lundgren A and Macas R 2023
  \href{https://doi.org/10.1103/PhysRevD.108.122004}{{\em Phys. Rev. D\/} {\bf
  108} 122004} (\textit{Preprint} \href{https://arxiv.org/abs/2309.06594}{{\tt
  2309.06594}})

\bibitem{Davis:2019}
Davis D, Massinger T~J, Lundgren A~P, Driggers J~C {\em et~al.\/} 2019
  \href{https://doi.org/10.1088/1361-6382/ab01c5}{{\em Class. Quant. Grav.\/}
  {\bf 36} 055011} (\textit{Preprint}
  \href{https://arxiv.org/abs/1809.05348}{{\tt 1809.05348}})

\bibitem{Cornish:2014kda}
Cornish N~J and Littenberg T~B 2015
  \href{https://doi.org/10.1088/0264-9381/32/13/135012}{{\em Class. Quant.
  Grav.\/} {\bf 32} 135012} (\textit{Preprint}
  \href{https://arxiv.org/abs/1410.3835}{{\tt 1410.3835}})

\bibitem{Pankow:2018qpo}
Pankow C {\em et~al.\/} 2018
  \href{https://doi.org/10.1103/PhysRevD.98.084016}{{\em Phys. Rev.\/} {\bf
  D98} 084016} (\textit{Preprint} \href{https://arxiv.org/abs/1808.03619}{{\tt
  1808.03619}})

\bibitem{Cornish:2020dwh}
Cornish N~J, Littenberg T~B, B\'ecsy B, Chatziioannou K {\em et~al.\/} 2021
  \href{https://doi.org/10.1103/PhysRevD.103.044006}{{\em Phys. Rev. D\/} {\bf
  103} 044006} (\textit{Preprint} \href{https://arxiv.org/abs/2011.09494}{{\tt
  2011.09494}})

\bibitem{Ashton:2022ztk}
Ashton G 2023 \href{https://doi.org/10.1093/mnras/stad341}{{\em Mon. Not. Roy.
  Astron. Soc.\/} {\bf 520} 2983--94} (\textit{Preprint}
  \href{https://arxiv.org/abs/2209.15547}{{\tt 2209.15547}})

\bibitem{Sesana:2009wg}
Sesana A, Gair J, Mandel I and Vecchio A 2009
  \href{https://doi.org/10.1088/0004-637X/698/2/L129}{{\em Astrophys. J.
  Lett.\/} {\bf 698} L129--32} (\textit{Preprint}
  \href{https://arxiv.org/abs/0903.4177}{{\tt 0903.4177}})

\bibitem{Carr:2020xqk}
Carr B and Kuhnel F 2020
  \href{https://doi.org/10.1146/annurev-nucl-050520-125911}{{\em Ann. Rev.
  Nucl. Part. Sci.\/} {\bf 70} 355--94} (\textit{Preprint}
  \href{https://arxiv.org/abs/2006.02838}{{\tt 2006.02838}})

\bibitem{Srivastava:2019fcb}
Srivastava V, Ballmer S, Brown D~A, Afle C {\em et~al.\/} 2019
  \href{https://doi.org/10.1103/PhysRevD.100.043026}{{\em Phys. Rev. D\/} {\bf
  100} 043026} (\textit{Preprint} \href{https://arxiv.org/abs/1906.00084}{{\tt
  1906.00084}})

\bibitem{Vartanyan:2023sxm}
Vartanyan D, Burrows A, Wang T, Coleman M~S~B and White C~J 2023
  \href{https://doi.org/10.1103/PhysRevD.107.103015}{{\em Phys. Rev. D\/} {\bf
  107} 103015} (\textit{Preprint} \href{https://arxiv.org/abs/2302.07092}{{\tt
  2302.07092}})

\bibitem{Pizzati:2021apa}
Pizzati E, Sachdev S, Gupta A and Sathyaprakash B 2022
  \href{https://doi.org/10.1103/PhysRevD.105.104016}{{\em Phys. Rev. D\/} {\bf
  105} 104016} (\textit{Preprint} \href{https://arxiv.org/abs/2102.07692}{{\tt
  2102.07692}})

\bibitem{Himemoto:2021ukb}
Himemoto Y, Nishizawa A and Taruya A 2021
  \href{https://doi.org/10.1103/PhysRevD.104.044010}{{\em Phys. Rev. D\/} {\bf
  104} 044010} (\textit{Preprint} \href{https://arxiv.org/abs/2103.14816}{{\tt
  2103.14816}})

\bibitem{Hu:2022bji}
Hu Q and Veitch J 2023 \href{https://doi.org/10.3847/1538-4357/acbc18}{{\em
  Astrophys. J.\/} {\bf 945} 103} (\textit{Preprint}
  \href{https://arxiv.org/abs/2210.04769}{{\tt 2210.04769}})

\bibitem{Kumar:2022tto}
Kumar S, Nitz A~H and Forteza X~J 2022 \textit{Preprint}
  \href{https://arxiv.org/abs/2202.12762}{{\tt 2202.12762}}

\bibitem{gstlal-methods}
Messick C {\em et~al.\/} 2017
  \href{https://doi.org/10.1103/PhysRevD.95.042001}{{\em Phys. Rev. D\/} {\bf
  95} 042001} (\textit{Preprint} \href{https://arxiv.org/abs/1604.04324}{{\tt
  1604.04324}})

\bibitem{Biswas:2019wmx}
Biswas A, McIver J and Mahabal A 2020
  \href{https://doi.org/10.1088/1361-6382/ab8650}{{\em Class. Quant. Grav.\/}
  {\bf 37} 175008} (\textit{Preprint}
  \href{https://arxiv.org/abs/1910.12143}{{\tt 1910.12143}})

\bibitem{Coughlin:2016lny}
Coughlin M {\em et~al.\/} 2017
  \href{https://doi.org/10.1088/1361-6382/aa5a60}{{\em Class. Quant. Grav.\/}
  {\bf 34} 044004} (\textit{Preprint}
  \href{https://arxiv.org/abs/1611.09812}{{\tt 1611.09812}})

\bibitem{Biscans:2017yce}
Biscans S {\em et~al.\/} 2018
  \href{https://doi.org/10.1088/1361-6382/aaa4aa}{{\em Class. Quant. Grav.\/}
  {\bf 35} 055004} (\textit{Preprint}
  \href{https://arxiv.org/abs/1707.03466}{{\tt 1707.03466}})

\bibitem{LIGO:2020pzq}
Schwartz E {\em et~al.\/} (LIGO Scientific) 2020
  \href{https://doi.org/10.1088/1361-6382/abbc8c}{{\em Class. Quant. Grav.\/}
  {\bf 37} 235007} (\textit{Preprint}
  \href{https://arxiv.org/abs/2007.12847}{{\tt 2007.12847}})

\bibitem{Colpi:2016fup}
Colpi M and Sesana A 2017 {\em {Gravitational Wave Sources in the Era of
  Multi-Band Gravitational Wave Astronomy}\/} pp 43--140 (\textit{Preprint}
  \href{https://arxiv.org/abs/1610.05309}{{\tt 1610.05309}})

\bibitem{Gerosa:2019dbe}
Gerosa D, Ma S, Wong K~W~K, Berti E {\em et~al.\/} 2019
  \href{https://doi.org/10.1103/PhysRevD.99.103004}{{\em Phys. Rev. D\/} {\bf
  99} 103004} (\textit{Preprint} \href{https://arxiv.org/abs/1902.00021}{{\tt
  1902.00021}})

\bibitem{Toubiana:2022vpp}
Toubiana A, Babak S, Marsat S and Ossokine S 2022
  \href{https://doi.org/10.1103/PhysRevD.106.104034}{{\em Phys. Rev. D\/} {\bf
  106} 104034} (\textit{Preprint} \href{https://arxiv.org/abs/2206.12439}{{\tt
  2206.12439}})

\bibitem{Fairhurst:2010is}
Fairhurst S 2011 \href{https://doi.org/10.1088/0264-9381/28/10/105021}{{\em
  Class. Quant. Grav.\/} {\bf 28} 105021} (\textit{Preprint}
  \href{https://arxiv.org/abs/1010.6192}{{\tt 1010.6192}})

\bibitem{T2100150}
Davis D {\em et~al.\/} 2021 {Data Quality Vetoes Applied to the Analysis of
  LIGO Data from the Second Observing Run} Tech. Rep. DCC-T2100150 {LIGO}
  \urlprefix\url{https://dcc.ligo.org/LIGO-T2100150/public}

\bibitem{T2100045}
Davis D {\em et~al.\/} 2021 {Data Quality Vetoes Applied to the Analysis of
  LIGO Data from the Third Observing Run} Tech. Rep. DCC-T2100045 {LIGO}
  \urlprefix\url{https://dcc.ligo.org/LIGO-T2100045/public}

\bibitem{Goetz:2023aaa}
Goetz E and Riles K 2023 {Segments used for creating standard SFTs in O3 data}
  Tech. Rep. DCC-T2300068 {LIGO}
  \urlprefix\url{https://dcc.ligo.org/LIGO-T2300068/public}

\bibitem{Goetz:2021aaa}
Goetz E {\em et~al.\/} 2021 {O3a lines and combs in found in self-gated C01
  data} Tech. Rep. DCC-T2100153 {LIGO}
  \urlprefix\url{https://dcc.ligo.org/LIGO-T2100153/public}

\bibitem{LSC:2018vzm}
Covas P~B {\em et~al.\/} (LSC) 2018
  \href{https://doi.org/10.1103/PhysRevD.97.082002}{{\em Phys. Rev. D\/} {\bf
  97} 082002} (\textit{Preprint} \href{https://arxiv.org/abs/1801.07204}{{\tt
  1801.07204}})

\bibitem{LIGOScientific:2020qhb}
Abbott R {\em et~al.\/} (LIGO Scientific, Virgo, VIRGO) 2021
  \href{https://doi.org/10.1103/PhysRevD.103.064017}{{\em Phys. Rev. D\/} {\bf
  103} 064017} [Erratum: Phys.Rev.D 108, 069901 (2023)] (\textit{Preprint}
  \href{https://arxiv.org/abs/2012.12128}{{\tt 2012.12128}})

\bibitem{LIGOScientific:2017csd}
Abbott B~P {\em et~al.\/} (LIGO Scientific, Virgo) 2017
  \href{https://doi.org/10.1103/PhysRevD.96.062002}{{\em Phys. Rev. D\/} {\bf
  96} 062002} (\textit{Preprint} \href{https://arxiv.org/abs/1707.02667}{{\tt
  1707.02667}})

\bibitem{Tenorio:2021wmz}
Tenorio R, Keitel D and Sintes A~M 2021
  \href{https://doi.org/10.3390/universe7120474}{{\em Universe\/} {\bf 7} 474}
  (\textit{Preprint} \href{https://arxiv.org/abs/2111.12575}{{\tt 2111.12575}})

\bibitem{Riles:2022wwz}
Riles K 2023 \href{https://doi.org/10.1007/s41114-023-00044-3}{{\em Living Rev.
  Rel.\/} {\bf 26} 3} (\textit{Preprint}
  \href{https://arxiv.org/abs/2206.06447}{{\tt 2206.06447}})

\bibitem{Wette:2023dom}
Wette K 2023 \href{https://doi.org/10.1016/j.astropartphys.2023.102880}{{\em
  Astropart. Phys.\/} {\bf 153} 102880} (\textit{Preprint}
  \href{https://arxiv.org/abs/2305.07106}{{\tt 2305.07106}})

\bibitem{Manchester:2004bp}
Manchester R~N, Hobbs G~B, Teoh A and Hobbs M 2005
  \href{https://doi.org/10.1086/428488}{{\em Astron. J.\/} {\bf 129} 1993}
  (\textit{Preprint} \href{https://arxiv.org/abs/astro-ph/0412641}{{\tt
  astro-ph/0412641}})

\bibitem{ATNF_catalog}
{ATNF} pulsar catalogue, v1.71
  \url{https://www.atnf.csiro.au/research/pulsar/psrcat/index.html?version=1.71}
  accessed: 2024-01-08

\bibitem{GWOSC_O3}
Abbott R {\em et~al.\/} (KAGRA, Virgo, LIGO Scientific) 2023
  \href{https://doi.org/10.3847/1538-4365/acdc9f}{{\em Astrophys. J. Suppl.\/}
  {\bf 267} 29} (\textit{Preprint} \href{https://arxiv.org/abs/2302.03676}{{\tt
  2302.03676}})

\bibitem{O3_lines}
{O3} instrumental lines \url{https://gwosc.org/O3/o3speclines/} accessed:
  2024-01-08

\bibitem{Tuyenbayev_2016}
Tuyenbayev D {\em et~al.\/} 2017
  \href{https://doi.org/10.1088/0264-9381/34/1/015002}{{\em Class. Quant.
  Grav.\/} {\bf 34} 015002} (\textit{Preprint}
  \href{https://arxiv.org/abs/1608.05134}{{\tt 1608.05134}})

\bibitem{Viets_2018}
Viets A {\em et~al.\/} 2018
  \href{https://doi.org/10.1088/1361-6382/aab658}{{\em Class. Quant. Grav.\/}
  {\bf 35} 095015} (\textit{Preprint}
  \href{https://arxiv.org/abs/1710.09973}{{\tt 1710.09973}})

\bibitem{alog:violin_lines}
Starkman T 2023 {aLIGO LHO Logbook}
  \href{https://alog.ligo-wa.caltech.edu/aLOG/index.php?callRep=71501}{71501}

\bibitem{robertson_BRD}
Robertson N~A, Fritschel P, Shapiro B, Torrie C~I and Evans M 2017
  \href{https://doi.org/10.1063/1.4978796}{{\em Rev. Sci. Instrum.\/} {\bf 88}
  035117}

\bibitem{Biscans_AMD}
Biscans S, Gras S, Blair C~D, Driggers J {\em et~al.\/} 2019
  \href{https://doi.org/10.1103/PhysRevD.100.122003}{{\em Phys. Rev. D\/} {\bf
  100} 122003} (\textit{Preprint} \href{https://arxiv.org/abs/1909.07805}{{\tt
  1909.07805}})

\bibitem{Driggers:2018gii}
Driggers J {\em et~al.\/} (LIGO Scientific) 2019
  \href{https://doi.org/10.1103/PhysRevD.99.042001}{{\em Phys. Rev. D\/} {\bf
  99} 042001} (\textit{Preprint} \href{https://arxiv.org/abs/1806.00532}{{\tt
  1806.00532}})

\bibitem{Vajente:2019ycy}
Vajente G, Huang Y, Isi M, Driggers J~C {\em et~al.\/} 2020
  \href{https://doi.org/10.1103/PhysRevD.101.042003}{{\em Phys. Rev. D\/} {\bf
  101} 042003} (\textit{Preprint} \href{https://arxiv.org/abs/1911.09083}{{\tt
  1911.09083}})

\bibitem{Viets:2021aaa}
Viets A and Wade M 2021 {Subtracting Narrow-band Noise from LIGO Strain Data in
  the Third Observing Run} Tech. Rep. DCC-T2100058 {LIGO}
  \urlprefix\url{https://dcc.ligo.org/LIGO-T2100058/public}

\bibitem{Yu:2021swq}
Yu H and Adhikari R~X 2022 \href{https://doi.org/10.3389/frai.2022.811563}{{\em
  Front. Artif. Intell.\/} {\bf 5} 811563} (\textit{Preprint}
  \href{https://arxiv.org/abs/2111.03295}{{\tt 2111.03295}})

\bibitem{Thrane:2014yza}
Thrane E, Christensen N, Schofield R~M~S and Effler A 2014
  \href{https://doi.org/10.1103/PhysRevD.90.023013}{{\em Phys. Rev. D\/} {\bf
  90} 023013} (\textit{Preprint} \href{https://arxiv.org/abs/1406.2367}{{\tt
  1406.2367}})

\bibitem{Coughlin:2016vor}
Coughlin M~W {\em et~al.\/} 2016
  \href{https://doi.org/10.1088/0264-9381/33/22/224003}{{\em Class. Quant.
  Grav.\/} {\bf 33} 224003} (\textit{Preprint}
  \href{https://arxiv.org/abs/1606.01011}{{\tt 1606.01011}})

\bibitem{Coughlin:2018tjc}
Coughlin M~W {\em et~al.\/} 2018
  \href{https://doi.org/10.1103/PhysRevD.97.102007}{{\em Phys. Rev. D\/} {\bf
  97} 102007} (\textit{Preprint} \href{https://arxiv.org/abs/1802.00885}{{\tt
  1802.00885}})

\bibitem{Meyers:2020qrb}
Meyers P~M, Martinovic K, Christensen N and Sakellariadou M 2020
  \href{https://doi.org/10.1103/PhysRevD.102.102005}{{\em Phys. Rev. D\/} {\bf
  102} 102005} (\textit{Preprint} \href{https://arxiv.org/abs/2008.00789}{{\tt
  2008.00789}})

\bibitem{Renzini:2022alw}
Renzini A~I, Goncharov B, Jenkins A~C and Meyers P~M 2022
  \href{https://doi.org/10.3390/galaxies10010034}{{\em Galaxies\/} {\bf 10} 34}
  (\textit{Preprint} \href{https://arxiv.org/abs/2202.00178}{{\tt 2202.00178}})

\bibitem{KAGRA:2021kbb}
Abbott R {\em et~al.\/} (KAGRA, Virgo, LIGO Scientific) 2021
  \href{https://doi.org/10.1103/PhysRevD.104.022004}{{\em Phys. Rev. D\/} {\bf
  104} 022004} (\textit{Preprint} \href{https://arxiv.org/abs/2101.12130}{{\tt
  2101.12130}})

\bibitem{GWTC3_pops}
Abbott R {\em et~al.\/} (KAGRA, Virgo, LIGO Scientific) 2023
  \href{https://doi.org/10.1103/PhysRevX.13.011048}{{\em Phys. Rev. X\/} {\bf
  13} 011048} (\textit{Preprint} \href{https://arxiv.org/abs/2111.03634}{{\tt
  2111.03634}})

\bibitem{ET_science}
Maggiore M {\em et~al.\/} 2020
  \href{https://doi.org/10.1088/1475-7516/2020/03/050}{{\em JCAP\/} {\bf 03}
  050} (\textit{Preprint} \href{https://arxiv.org/abs/1912.02622}{{\tt
  1912.02622}})

\bibitem{Janssens:2021cta}
Janssens K, Martinovic K, Christensen N, Meyers P~M and Sakellariadou M 2021
  \href{https://doi.org/10.1103/PhysRevD.104.122006}{{\em Phys. Rev. D\/} {\bf
  104} 122006} [Erratum: Phys.Rev.D 105, 109904 (2022)] (\textit{Preprint}
  \href{https://arxiv.org/abs/2110.14730}{{\tt 2110.14730}})

\bibitem{Janssens:2022xmo}
Janssens K, Boileau G, Christensen N, Badaracco F and van Remortel N 2022
  \href{https://doi.org/10.1103/PhysRevD.106.042008}{{\em Phys. Rev. D\/} {\bf
  106} 042008} (\textit{Preprint} \href{https://arxiv.org/abs/2206.06809}{{\tt
  2206.06809}})

\bibitem{Wong:2021eun}
Wong I~C~F and Li T~G~F 2022
  \href{https://doi.org/10.1103/PhysRevD.105.084002}{{\em Phys. Rev. D\/} {\bf
  105} 084002} (\textit{Preprint} \href{https://arxiv.org/abs/2108.05108}{{\tt
  2108.05108}})

\bibitem{Janssens:2022cty}
Janssens K, Boileau G, Bizouard M~A, Christensen N {\em et~al.\/} 2023
  \href{https://doi.org/10.1140/epjp/s13360-023-03948-9}{{\em Eur. Phys. J.
  Plus\/} {\bf 138} 352} [Erratum: Eur.Phys.J.Plus 138, 446 (2023)]
  (\textit{Preprint} \href{https://arxiv.org/abs/2205.00416}{{\tt 2205.00416}})

\bibitem{Cireddu:2023ssf}
Cireddu F, Wils M, Wong I~C~F, Pang P~T~H {\em et~al.\/} 2023 \textit{Preprint}
  \href{https://arxiv.org/abs/2312.14614}{{\tt 2312.14614}}

\bibitem{Lindblom:2009un}
Lindblom L 2009 \href{https://doi.org/10.1103/PhysRevD.80.042005}{{\em Phys.
  Rev. D\/} {\bf 80} 042005} (\textit{Preprint}
  \href{https://arxiv.org/abs/0906.5153}{{\tt 0906.5153}})

\bibitem{sun2021characterization}
Sun L {\em et~al.\/} 2021 \textit{Preprint}
  \href{https://arxiv.org/abs/2107.00129}{{\tt 2107.00129}}

\bibitem{Virgo:2021kfv}
Acernese F {\em et~al.\/} (Virgo) 2022
  \href{https://doi.org/10.1088/1361-6382/ac3c8e}{{\em Class. Quant. Grav.\/}
  {\bf 39} 045006} (\textit{Preprint}
  \href{https://arxiv.org/abs/2107.03294}{{\tt 2107.03294}})

\bibitem{Vitale:2011wu}
Vitale S, Del~Pozzo W, Li T~G~F, Van Den~Broeck C {\em et~al.\/} 2012
  \href{https://doi.org/10.1103/PhysRevD.85.064034}{{\em Phys. Rev.\/} {\bf
  D85} 064034} (\textit{Preprint} \href{https://arxiv.org/abs/1111.3044}{{\tt
  1111.3044}})

\bibitem{Hall_2019}
Hall E~D, Cahillane C, Izumi K, Smith R~J~E and Adhikari R~X 2019
  \href{https://doi.org/10.1088/1361-6382/ab368c}{{\em Class. Quant. Grav.\/}
  {\bf 36} 205006} (\textit{Preprint}
  \href{https://arxiv.org/abs/1712.09719}{{\tt 1712.09719}})

\bibitem{Payne_2020}
Payne E, Talbot C, Lasky P~D, Thrane E and Kissel J~S 2020
  \href{https://doi.org/10.1103/PhysRevD.102.122004}{{\em Phys. Rev. D\/} {\bf
  102} 122004} (\textit{Preprint} \href{https://arxiv.org/abs/2009.10193}{{\tt
  2009.10193}})

\bibitem{Vitale_2021}
Vitale S, Haster C~J, Sun L, Farr B {\em et~al.\/} 2021
  \href{https://doi.org/10.1103/physrevd.103.063016}{{\em Physical Review D\/}
  {\bf 103}} \urlprefix\url{http://dx.doi.org/10.1103/physrevd.103.063016}

\bibitem{Huang:2022rdg}
Huang Y, Chen H~Y, Haster C~J, Sun L {\em et~al.\/} 2022 \textit{Preprint}
  \href{https://arxiv.org/abs/2204.03614}{{\tt 2204.03614}}

\bibitem{Wade_2023}
Wade M, Viets A~D, Chmiel T, Stover M and Wade L 2023
  \href{https://doi.org/10.1088/1361-6382/acabf6}{{\em Classical and Quantum
  Gravity\/} {\bf 40} 035001}
  \urlprefix\url{http://dx.doi.org/10.1088/1361-6382/acabf6}

\bibitem{Read_2023}
Read J 2023 \href{https://doi.org/10.1088/1361-6382/acd29b}{{\em Classical and
  Quantum Gravity\/} {\bf 40} 135002}
  \urlprefix\url{https://dx.doi.org/10.1088/1361-6382/acd29b}

\bibitem{Chen_2021}
Chen H~Y, Cowperthwaite P~S, Metzger B~D and Berger E 2021
  \href{https://doi.org/10.3847/2041-8213/abdab0}{{\em Astrophys. J. Lett.\/}
  {\bf 908} L4} (\textit{Preprint} \href{https://arxiv.org/abs/2011.01211}{{\tt
  2011.01211}})

\bibitem{Schutz_2018}
Schutz B~F 2018 \href{https://doi.org/10.1098/rsta.2017.0279}{{\em Phil. Trans.
  Roy. Soc. Lond. A\/} {\bf 376} 20170279} (\textit{Preprint}
  \href{https://arxiv.org/abs/1804.06308}{{\tt 1804.06308}})

\bibitem{Hernandez_Vivanco_2019}
Hernandez~Vivanco F, Smith R, Thrane E, Lasky P~D {\em et~al.\/} 2019
  \href{https://doi.org/10.1103/PhysRevD.100.103009}{{\em Phys. Rev. D\/} {\bf
  100} 103009} (\textit{Preprint} \href{https://arxiv.org/abs/1909.02698}{{\tt
  1909.02698}})

\bibitem{Chatziioannou_2020}
Chatziioannou K 2020 \href{https://doi.org/10.1007/s10714-020-02754-3}{{\em
  General Relativity and Gravitation\/} {\bf 52}}
  \urlprefix\url{http://dx.doi.org/10.1007/s10714-020-02754-3}

\bibitem{MIZUNO1993273}
Mizuno J, Strain K~A, Nelson P~G, Chen J~M {\em et~al.\/} 1993
  \href{https://doi.org/10.1016/0375-9601(93)90620-F}{{\em Phys. Lett. A\/}
  {\bf 175} 273--6}

\bibitem{PhysRevD.65.042001}
Buonanno A and Chen Y 2002
  \href{https://doi.org/10.1103/PhysRevD.65.042001}{{\em Phys. Rev. D\/} {\bf
  65} 042001} (\textit{Preprint}
  \href{https://arxiv.org/abs/gr-qc/0107021}{{\tt gr-qc/0107021}})

\bibitem{Ward:thesis}
Ward R~L 2010 {\em Length Sensing and Control of a Prototype Advanced
  Interferometric Gravitational Wave Detector\/} Ph.D. thesis Caltech Pasadena,
  CA

\bibitem{Hall:thesis}
Hall E 2017 {\em Long-Baseline Laser Interferometry for the Detection of Binary
  Black-Hole Mergers\/} Ph.D. thesis Caltech Pasadena, CA

\bibitem{LIGO:DCC-T050136}
Elliot H 2005 {Analysis of the Frequency Dependence of LIGO Directional
  Sensitivity (Antenna Pattern) and Implications for Detector Calibration}
  Tech. Rep. DCC-T050136 {LIGO}
  \urlprefix\url{https://dcc.ligo.org/LIGO-T050136/public}

\bibitem{martynov_2019}
Martynov D {\em et~al.\/} 2019
  \href{https://doi.org/10.1103/PhysRevD.99.102004}{{\em Phys. Rev. D\/} {\bf
  99} 102004} (\textit{Preprint} \href{https://arxiv.org/abs/1901.03885}{{\tt
  1901.03885}})

\bibitem{Rakhmanov_2008}
Rakhmanov M, Romano J~D and Whelan J~T 2008
  \href{https://doi.org/10.1088/0264-9381/25/18/184017}{{\em Class. Quant.
  Grav.\/} {\bf 25} 184017} (\textit{Preprint}
  \href{https://arxiv.org/abs/0808.3805}{{\tt 0808.3805}})

\bibitem{Essick_2017}
Essick R, Vitale S and Evans M 2017
  \href{https://doi.org/10.1103/physrevd.96.084004}{{\em Physical Review D\/}
  {\bf 96}} \urlprefix\url{http://dx.doi.org/10.1103/physrevd.96.084004}

\bibitem{LIGO:DCC-T060237}
Rakhmanov M 2006 {Response of LIGO to Gravitational Waves at High Frequencies
  and in the Vicinity of the FSR} Tech. Rep. DCC-T060237 {LIGO}
  \urlprefix\url{https://dcc.ligo.org/LIGO-T060237/public}

\bibitem{Rakhmanov_2002}
Rakhmanov M, Savage R, Reitze D and Tanner D 2002
  \href{https://doi.org/10.1016/s0375-9601(02)01469-x}{{\em Phys. Lett. A\/}
  {\bf 305} 239–244}

\bibitem{LIGO:DCC-T970101}
Sigg D 1997 {Strain Calibration in LIGO} Tech. Rep. DCC-T970101 {LIGO}
  \urlprefix\url{https://dcc.ligo.org/LIGO-T970101/public}

\bibitem{Goetz_2009}
Goetz E {\em et~al.\/} 2009
  \href{https://doi.org/10.1088/0264-9381/26/24/245011}{{\em Class. Quant.
  Grav.\/} {\bf 26} 245011} (\textit{Preprint}
  \href{https://arxiv.org/abs/0910.5591}{{\tt 0910.5591}})

\bibitem{Goetz:2009wf}
Goetz E {\em et~al.\/} 2010
  \href{https://doi.org/10.1088/0264-9381/27/8/084024}{{\em Class. Quant.
  Grav.\/} {\bf 27} 084024} (\textit{Preprint}
  \href{https://arxiv.org/abs/0911.0853}{{\tt 0911.0853}})

\bibitem{Hild_2007}
Hild S {\em et~al.\/} 2007
  \href{https://doi.org/10.1088/0264-9381/24/22/025}{{\em Class. Quant.
  Grav.\/} {\bf 24} 5681--8} (\textit{Preprint}
  \href{https://arxiv.org/abs/0710.1229}{{\tt 0710.1229}})

\bibitem{Estevez_2018}
Estevez D, Lieunard B, Marion F, Mours B {\em et~al.\/} 2018
  \href{https://doi.org/10.1088/1361-6382/aae95f}{{\em Class. Quant. Grav.\/}
  {\bf 35} 235009} (\textit{Preprint}
  \href{https://arxiv.org/abs/1806.06572}{{\tt 1806.06572}})

\bibitem{Estevez_2021}
Estevez D, Mours B and Pradier T 2021
  \href{https://doi.org/10.1088/1361-6382/abe2da}{{\em Class. Quant. Grav.\/}
  {\bf 38} 075012} (\textit{Preprint}
  \href{https://arxiv.org/abs/2011.03728}{{\tt 2011.03728}})

\bibitem{Ross_2021}
Ross M~P {\em et~al.\/} 2021
  \href{https://doi.org/10.1103/PhysRevD.104.082006}{{\em Phys. Rev. D\/} {\bf
  104} 082006} (\textit{Preprint} \href{https://arxiv.org/abs/2107.00141}{{\tt
  2107.00141}})

\bibitem{Inoue_2018}
Inoue Y, Haino S, Kanda N, Ogawa Y {\em et~al.\/} 2018
  \href{https://doi.org/10.1103/PhysRevD.98.022005}{{\em Phys. Rev. D\/} {\bf
  98} 022005} (\textit{Preprint} \href{https://arxiv.org/abs/1804.08249}{{\tt
  1804.08249}})

\bibitem{Goetz_2010}
Goetz E and Savage R~L 2010
  \href{https://doi.org/10.1088/0264-9381/27/21/215001}{{\em Classical and
  Quantum Gravity\/} {\bf 27} 215001}
  \urlprefix\url{https://dx.doi.org/10.1088/0264-9381/27/21/215001}

\bibitem{Essick_2019}
Essick R and Holz D~E 2019 \href{https://doi.org/10.1088/1361-6382/ab2142}{{\em
  Class. Quant. Grav.\/} {\bf 36} 125002} (\textit{Preprint}
  \href{https://arxiv.org/abs/1902.08076}{{\tt 1902.08076}})

\bibitem{schutz2020selfcalibration}
Schutz B~F and Sathyaprakash B~S 2020 Self-calibration of networks of
  gravitational wave detectors (\textit{Preprint}
  \href{https://arxiv.org/abs/2009.10212}{{\tt 2009.10212}})

\bibitem{LIGOScientific:2007fwp}
Abbott B~P {\em et~al.\/} (LIGO Scientific) 2009
  \href{https://doi.org/10.1088/0034-4885/72/7/076901}{{\em Rept. Prog.
  Phys.\/} {\bf 72} 076901} (\textit{Preprint}
  \href{https://arxiv.org/abs/0711.3041}{{\tt 0711.3041}})

\bibitem{Bartos:2011aaa}
Bartos I {\em et~al.\/} 2011 {Frequency Domain Calibration Error Budget for
  LIGO in S6} Tech. Rep. DCC-T1100071 {LIGO}
  \urlprefix\url{https://dcc.ligo.org/LIGO-T1100071/public}

\bibitem{GWTC3_figures}
{LIGO-Virgo-KAGRA Collaboration} 2023 {GWTC-3: Compact Binary Coalescences
  Observed by LIGO and Virgo During the Second Part of the Third Observing Run
  — Data behind the figures}
  \urlprefix\url{https://doi.org/10.5281/zenodo.7997424}

\bibitem{GWTC3_search_sens}
{LIGO-Virgo-KAGRA Collaboration} 2023 {GWTC-3: Compact Binary Coalescences
  Observed by LIGO and Virgo During the Second Part of the Third Observing Run
  — O3 search sensitivity estimates}
  \urlprefix\url{https://doi.org/10.5281/zenodo.7890437}

\bibitem{Pearson_corr}
{Pearson} K 1895 {\em Proceedings of the Royal Society of London Series I\/}
  {\bf 58} 240--2

\end{thebibliography}

\end{document}